\documentclass{JHEP3}

\usepackage{amsfonts,amsbsy,amsmath,amssymb}
\usepackage{epsfig}
\usepackage{subfigure}
\newcommand{\thetachar}[2]{\vartheta\left[ \begin{array}{c}
          #1 \\ #2
                    \end{array}\right]}

\newcommand{\be}{\begin{equation}}
\newcommand{\ee}{\end{equation}}
\newcommand{\bea}{\begin{eqnarray}}
\newcommand{\eea}{\end{eqnarray}}

\title{Dynamics of critical vortices  on the torus and on the plane}

\author{Antonio Gonz\'alez-Arroyo and Alberto Ramos\\ Departamento de
  F\'{\i}sica Te\'orica Universidad Aut\'onoma de   Madrid and\\
    Instituto de F\'{\i}sica Te\'orica UAM-CSIC
    Universidad Aut\'onoma de Madrid \\ Cantoblanco
    Madrid 28049 SPAIN     \\ E-mail:
\href{mailto:antonio.gonzalez-arroyo@uam.es}{\email{antonio.gonzalez-arroyo@uam.es}},
\href{mailto:alberto@martin.ft.uam.es}{\email{alberto@martin.ft.uam.es}}
}

\abstract{
We use the Bradlow parameter expansion to construct the metric tensor in the 
space of solutions of the Bogomolny equations for the Abelian Higgs model on a
two-dimensional torus. Using this metric we study  the dynamics and
scattering of vortices on the torus within the geodesic approximation.
For small torus volumes the metric is determined in terms of a small 
number of parameters. For large volumes the results provide a very
precise approximation to the metric and dynamics on the plane.
}

\preprint{\hepth{XXXXXX}; FTUAM-05-19; IFT-UAM/CSIC-05-53}

\maketitle

\begin{document}

\section{Introduction}

Topological defects are interesting classically stable structures
which are present in many field theories. Their stability has a
topological origin. This explains why they are relevant 
for many areas  of Physics. One of the most fascinating class of 
defects is given by Abrikosov vortices~\cite{Abrikosov:1956sx}. In
three-dimensional space they 
arise as one-dimensional (string-like) structures  acting as magnetic
flux-tubes in type II superconductors.  In a  simplified  setting, they are
minimum energy configurations in a two-dimensional  Abelian-Higgs
gauge model~\cite{no:vortice}. Their stability arises from magnetic flux
conservation.  Apart from Superconductivity in Condensed Matter Physics
there are many areas of Physics where this type of solutions might be
relevant. One of the most fascinating is in Cosmology, in which they have
been studied in connection to  Cosmic strings for quite a
while~\cite{Kibble:1976sj,Vilenkin:1981kz,Vilenkin:1994sh}.
There seems to be a wide class of possible extensions of the Standard
Model~\cite{Kibble:2004hq,Jeannerot:2003} which possess this kind of
solutions. Furthermore, other extensions involving extra spatial
compact dimensions do provide additional situations in which vortex
solutions can appear and become
relevant~\cite{Libanov:2000uf,Frere:2000dc,Frere:2003yv}. They have also
been studied within the standard model itself in connection with the
dual-superconductivity scenario of Confinement.

Being interesting mathematical objects in their own right and having shown a
factual and/or potential interest in Physics, vortex-solutions  have
been studied for a long time~\cite{Jaffe:1980ja}. A recent
book~\cite{Manton:2004tk} 
compiles many results accumulated over the years. The most fascinating
situation occurs for a particular value of the Higgs self-coupling marking
the transition from type-I to type-II superconductors. In this case,
known as the  
Bogomolny limit~\cite{Bogomolny:1976de}, the field equations (expressing the
condition of minimal potential energy per unit length of the string) reduce
to the first order Bogomolny equations. Taubes~\cite{Taubes:1980tm}
has demonstrated that 
given $q$ points (the location of the zeros of the Higgs field)
there exist a unique solution to the Bogomolny equations (modulo gauge
transformations). Despite this simplification
there is no analytical expression for the solutions. For
spherically(in 2-d) symmetric solutions, including the single vortex
case, the asymptotic behaviour and a  Taylor expansion in the distance
to the centre~\cite{deVega:1976mi} has been obtained.

In Ref.~\cite{Gonzalez-Arroyo:2004xu} we proposed a method to
construct the solutions on the torus by means of an analytic expansion
in a  parameter measuring the departure of the area from the critical
value (twice the flux, in units of the Higgs mass). It is known that for
smaller areas there are no solutions of the Bogomolny equations. The
critical value of the area is the only case for which there is an
explicit analytic solution of the Bogomolny equations, although a fairly dull
one (vanishing Higgs field and constant magnetic field). In 
Ref.~\cite{Gonzalez-Arroyo:2004xu}
we computed the first 51 terms in the expansion and found the method
capable of describing the profile of the solution even for fairly
large torus areas, and arbitrary positions of the Higgs zeros.

We should stress here that, as mentioned in
Ref.~\cite{Gonzalez-Arroyo:2004xu},  the usefulness  
of the idea goes much  beyond the torus case. Indeed,
Bradlow~\cite{Bradlow:1990ir} 
discovered that the Bogomolny equations can be generalised to a class of
gauge-Higgs systems in compact K\"ahler manifolds. In all cases there is a
critical volume and  the pattern is repeated. This critical case is
referred as the {\em Bradlow limit}, and corresponds to a particular value of
a parameter appearing in Bradlow's formulation related to the volume. 
Our approach is that of expanding around this value of the {\em Bradlow
parameter} and is easily generalisable to other K\"ahler manifolds,
besides the torus. More general instances of this approach for higher
dimensional theories are also possible. As a matter of fact, our
proposal originated from the 
study of self-dual non-abelian gauge fields on the torus. In that case there
are also no known analytic solutions to the self-duality equations except 
for  constant field strength solutions occurring for particular sizes of the
torus. In Ref.~\cite{GarciaPerez:2000yt} one of the present authors and
collaborators showed that one could consistently expand the solution for
arbitrary  sizes in a power series in a parameter measuring the
departure from the critical aspect ratios (which allow for analytical
solutions). The general setting, to be addressed
elsewhere~\cite{Tony:pr} involves perturbations of the metric tensor
about the particular values allowing for constant(uniform) 
solutions. Finally, we should also comment that our method has also proven
useful for the study of non-compact manifolds. For example, if one is
interested in studying vortices on the two-dimensional plane then the torus
can be considered as an infrared cut-off to be removed in the infinite
area limit
limit. Indeed, our results of the previous paper and the 
present one show that this methodology is competitive with other
alternatives. 

One of the advantages of having an analytic method is that
it can be generalised with fairly modest effort to other questions involving 
vortices. The present paper is devoted to one of the most interesting
applications which is that of studying the scattering of vortices. Previously, 
numerical methods had been employed to study this
point~\cite{Samols:1990kd,Samols:1991ne,Moriarty:1988fx}. Indeed, what
we will do is to make use of our expansion 
to compute the metric of the manifold of solutions of the Bogomolny
equations on the torus. This leads, within the geodesic
approximation~\cite{Manton:1981mp}, to the determination of the classical
dynamics of these vortices.  The next section will be devoted to the
presentation of the  formalism and derivation of our main formulas. 
In particular, we derive the general structure of the metric tensor 
to any order, as a polynomial in certain coordinates of the moduli space. 
The explicit computation  of the first two orders for arbitrary number of vortices 
are collected in the Appendix. We then focused in the particular case 
of two-vortex scattering. The corresponding results are presented 
in  Section~\ref{sc:results}. In particular, we give the value
(up to machine double precision) of the coefficients 
of the polynomial which defines the metric up to fifth order in the 
expansion.  This should provide a very good
approximation to the metric not too far from the Bradlow limit. 
As we approach the infinite area limit more and more terms of the
expansion are required. We went up to order 40 in our expansion to compute
the salient features of the scattering of vortices on the plane. Here,
instead of computing the coefficients of the polynomials,  we defined a grid
in the moduli space and computed the metric at these points (up to
order 40). From these values we determined the metric tensor for
vortices in the plane with an accuracy better  than one per mille. 
Finally, in the last section of the  paper we give a summary and conclusions.

\section{Critical vortex dynamics in the geodesic approximation}

\subsection{Vortices and the Geodesic approximation}
\label{sc:moduli}
Let us begin by recalling the Lagrangian density of the abelian
Higgs model in four-dimensional($4D$) Minkowski space-time:
\begin{equation}
  \mathcal L =
  -\frac{1}{4}F_{\mu\nu}F^{\mu\nu} +
  \frac{1}{2}\overline{(D_\mu\phi)}(D^\mu\phi) 
  - \frac{\lambda}{8}(|\phi|^2-1)^2
\end{equation}
where $\phi$ is a complex scalar field, and $D_\mu = \partial_\mu -
\imath A_\mu$ is the covariant derivative with respect to $A_\mu$, the $U(1)$ gauge
potential. There exist stationary, z-independent   solutions of the classical 
equations of motion characterised by a quantised magnetic flux $2\pi
q$ ($q$ being an integer) through the x-y plane, known as multi-vortex
solutions. The Nielsen-Olesen~\cite{no:vortice} vortex (N-O vortex) is the simplest
solution, having $q=1$. Its  potential energy distribution has the
shape of a thick string parallel to the z-axis having a finite string
tension (energy per unit length).  A particularly interesting case,
which we will refer as ``critical'',  occurs for $\lambda=1$. In that
case the equations of motion reduce to first order equations, known as
Bogomolny equations, that in our units and for positive flux are given
by: 
\begin{subequations}
    \label{Bogomolny_eqs}
    \begin{equation}
      (D_1+\imath D_2)\phi = 0
    \end{equation}
    \begin{equation}
      B = \frac{1}{2}(1-|\phi|^2)
    \end{equation}
\end{subequations}
where $B$ is the $z$-component of the magnetic field.
The corresponding solutions, for which the string tension is now given
exactly given by  $\pi |q|$, will be referred as critical multi-vortex
solutions (in the  literature they are  sometimes called {\em self-dual}
vortices).  The space of multi-vortex
solutions is continuous and can be  parameterised by $2q$ real numbers,
which can be naively  interpreted as the the x-y locations of $q$ N-O
critical vortices. The potential energy profile for widely separated vortices
supports this interpretation. Individual N-O  vortices are exponentially 
localised (energy density decreases exponentially
away from its centre) with a typical size of order $1$ in our units. For
vortex separations of this order and smaller, the field distribution
begins to depart sizably from a naive superposition of N-O vortices, but still
the Higgs field vanishes at $q$ locations in the x-y
plane~\cite{Taubes:1980tm}. The location of these zeroes will hence
serve as coordinates  labelling the space of multi-vortices. 

If we are interested in dynamical processes, we can consider solutions of the
classical equations of motion which tend to well-separated vortices in the 
past. This can be regarded as a scattering process involving  two or
more vortices. 
Numerical experiments have been performed by integrating the equations of
motion to describe this evolution 
process~\cite{Moriarty:1988fx,Shellard:1988zx,Samols:1991ne}. For
small initial velocities of the vortices it was
suggested~\cite{Manton:1981mp} that the  motion can be well-approximated by
the motion within the space of multi-vortex solutions, as this costs
no potential energy. This is referred as {\em geodesic
  motion}~\cite{Manton:1981mp,Ruback:1988ba}.  This is confirmed to be
rigorous in the limit of small velocities~\cite{Stuart:1994tc}, and 
numerically~\cite{Samols:1991ne}  it has been shown to be a good
approximation even up to relatively high velocities ($0.4c$).   

Within the context of the geodesic approximation,  the kinetic term in
the action becomes a quadratic form in the time derivatives of the
position of the q zeros of the Higgs field. From a geometrical
stand-point this quadratic form can be regarded as a  metric in
multi-vortex space. The  trajectories which are solutions of the
equations of motion are  the  geodesics corresponding to this metric.
This explains the name given to this approximation. 

Notice that the non-trivial scattering of vortices destroys the naive
interpretation of vortices as non-interacting objects. The latter 
image is based on the fact that the potential energy is 
independent of the position of the vortices. 
Instead, a very nice physical interpretation of vortices can be
given~\cite{Manton:2002wb} in which a vortex is considered a point
particle with a magnetic charge and a magnetic dipole moment. With
this in mind, the interaction of two static vortices can be shown to
be zero, but when the vortices move along some trajectory, it can be
shown that the interaction between them is exactly what is expected
for large separation between vortex centres.

Despite the nice formalism and results described in the previous paragraphs,
the main difficulty in studying both static and dynamical properties of
vortices is that there are no known analytical solutions of the Bogomolny
equations, not  even for the $q=1$ case. Most studies rely on numerical
methods  which often make use of symmetries to reduce
complexity~\cite{no:vortice,deVega:1976mi,rb:vortice}.

\subsection{Vortices in compact manifolds}

One can extend  the previous results to more general space-times. For
example, one can generalise the Bogomolny
equations~\cite{Bradlow:1990ir} to a class of systems with Higgs and
gauge fields in spaces  which are arbitrary K\"ahler manifolds. In
this construction there is  a  natural parameter, which we will call
{\em the Bradlow parameter},  which measures the spatial volume  in
units of the inverse photon or Higgs mass (which are equal at
criticality). In what follows, we will restrict to 2 dimensional
compact spaces, extending  the previously described case of Euclidean
space. These solutions can be regarded as solutions in higher 
dimensional spaces, not depending on the additional coordinates.  The
space of solutions (moduli space) of the corresponding  Bogomolny equations
turn out to be $2q$-dimensional manifolds. The kinetic term induces
a metric and, hence, a Riemannian structure on them. One can also
introduce a complex structure and show~\cite{Samols:1991ne} that they
become K\"ahler manifolds. Some global properties, such as the total
volume of these manifolds of moduli~\cite{Manton:1998kq}, can be
determined in terms of the properties of the 2-d space manifold.   

Obviously, the simplest (and usually the most interesting) cases  arise for the
sphere $S^2$ and the 2-torus  $\mathbb T^2$, which have been  explicitly
addressed in the
literature~\cite{Baptista:2002kb,Shah:1994us,Manton:1993tt}. The
interest of these 
extensions goes beyond the simply academic or purely mathematical
one. The torus might, for example, represent the behaviour of vortices
on a dense environment. Furthermore, there is an increasing interest
in vortex solutions within physical scenarios with extra
dimensions~\cite{Libanov:2000uf,Frere:2000dc,Frere:2003yv}. Frequently
the latter are compact and have two-dimensional cycles, so that these
types of solutions are present. A final motivation could follow from
our work, which shows that one can use the torus results to obtain
information about vortices on the plane. From now on we  will
concentrate on the flat torus case. 

Let us consider a two dimensional torus, described as a single patch with 
coordinates $x_1$ and $x_2$ having orthogonal periods of lengths $l_1$ and
$l_2$ respectively. We consider the  metric to be Euclidean
(other constant metrics can be easily considered), so that the total
area is given by $\mathcal A = l_1l_2$. We can also introduce the
complex coordinate $z$  and its complex conjugate $\overline z$ (in
general, we will denote complex conjugation  with an ``overline''
symbol), and their corresponding partial derivatives: 
\begin{subequations}
  \begin{equation}
    z = \frac{x_1+\imath x_2}{2};\qquad
    \overline z = \frac{x_1-\imath x_2}{2}
  \end{equation}
  \begin{equation}
    \partial = \partial_1 - \imath \partial_2;\qquad
    \overline\partial=\partial_1+\imath \partial_2
  \end{equation}
\end{subequations}
Now we will briefly revise
our formulation, as given in Ref.~\cite{Gonzalez-Arroyo:2004xu}.
The vector potential (as any  one-form in this complex manifold) can be
combined into a complex valued function ($(1,0)$-form)
$A(z,\overline z) = A_1(x_i) - \imath A_2(x_i)$.
However, this is not necessarily periodic, as a one-form should. Its
periodicity properties depend on the transition functions on the bundle
(See Ref.~\cite{ga:torus} for a simple introduction to the subject of gauge
fields on the torus). However, if $A^{(0)}$  is a specific  gauge field 
satisfying the boundary conditions, then $A-A^{(0)}$ is periodic and we can 
make use of the Hodge decomposition theorem to parametrise our gauge field
as:
\begin{equation}
\label{parametrization}
  A = - \imath \partial \overline H + v + A^{(0)}
\end{equation}
where $H=h-\imath g$ is a complex-valued  periodic  function of the variables
$z,\overline z$, and $v$ is a complex constant labelling the space of
harmonic forms on the torus. The function $g$ describes periodic gauge
transformations which are continuously connected with the identity. 
The remaining discrete set of gauge transformations imply that one can 
regard $v$ as the complex coordinate of the dual-torus (having lengths
$\frac{2\pi}{l_1},\frac{2\pi}{l_2}$). 
The periodicity in $v$ can be easily guessed by computing the 
corresponding Polyakov loops (the holonomy). 

Although a good deal of our formulas do not require a choice of the 
transition functions, for actual calculations one is required. 
In that case we will take  the following quasi-periodicity conditions 
for the  field (section of the bundle):
\begin{eqnarray}
  \label{period_cond}
    \phi(x_1+l_1,x_2) & = & e^{ \imath \pi
      q\frac{x_2}{l_2}}\phi(x_1,x_2) \\
    \nonumber
    \phi(x_1,x_2+l_2) & = & e^{-\imath \pi q\frac{x_1}{l_1}}\phi(x_1,x_2) 
\end{eqnarray}
where $q$ is the first Chern-number of the bundle.

There is a natural choice for $A^{(0)}$ which we will adopt, namely that 
which leads to a uniform field strength, $B^{(0)}\equiv f= \frac{2\pi
  q}{\mathcal A}$. With our choice of the transition functions
$A^{(0)}$ is given by
\begin{equation}
  A^{(0)} = -\imath f\overline z
\end{equation}
We will complement our parametrisation of the gauge field
Eq.~\ref{parametrization}, with the following one for the Higgs field:
\begin{equation}
  \phi=\sqrt{\epsilon} e^{-H} \chi
\end{equation}
where $\epsilon = 1-2f$. The function $\chi$ satisfies the same boundary
conditions  as the Higgs field (Eq.~\ref{period_cond} for our choice of
transition functions). 

Finally, we can write down the Bogomolny equations in our
parametrisation: 
\begin{subequations}
 \label{BOGBOTH}
  \begin{equation}
    \label{BOGONE}
    (\overline\partial +fz)\chi = \imath  \overline{v} \chi
  \end{equation}
  \begin{equation}
    \label{BOGTWO}
    \partial\overline\partial h = \frac{\epsilon}{2}\left(1 - e^{-2h}|\chi|^2
    \right) 
  \end{equation}
\end{subequations}
These equations can be solved sequentially. First, one solves for $\chi$ using
the first equation.  Plugging the result into the second equation (known as
the vortex equation) one solves for $h$. As we will show, the solution 
of the second equation is unique given $\chi$. Thus,  the whole structure of 
the moduli space of multi-vortex solutions resides in the first
equation. 

The constant $v$ can be taken as one of the coordinates of the moduli
space. The remaining coordinates can be associated to the space of
solutions at $v=0$. This is so because given a solution $\chi^{(v)}$
of Eq.~\ref{BOGONE} for some value of $v\neq 0$ one can associate to
it a unique solution $\chi^{(0)}$ for $v=0$ as follows
\begin{equation}
\chi^{(0)}(z,\overline z) = \mathcal T_{v} \left[\chi^{(v)}(z,\overline z)\right] \equiv
  e^{-\frac{|v|^2}{4f}}e^{\frac{-\imath }{2}(vz+\overline v\overline z)} 
  \chi^{(v)}\left(z+\frac{\imath \overline v}{2f}, \overline z -
    \frac{\imath v}{2f}\right)
\end{equation}
Which is a translation up to a gauge transformation and a rescaling by
a constant factor. This result extends to Eq.~\ref{BOGTWO} by also
translating  the function $h$, and adding  a
constant to it. This shows that the moduli space of solutions has the 
structure of a fibre bundle whose base space is a torus parametrised
by $v$. 

As explained in Ref.~\cite{Gonzalez-Arroyo:2004xu}, for $v=0$ 
Eq.~\ref{BOGONE} has $|q|$ linearly independent solutions,
which we will label $\chi_i^{(0)}(x)$. In that paper several choices of 
$\chi_i^{(0)}(x)$ are given. In particular, we will employ the
following one\footnote{\label{ft:theta}The symbol
  $\thetachar{a}{b}(z|t)$ 
  represent the theta function with characteristics $(a,b)$,
  and is
  given by
  \begin{equation}
    \vartheta\left[ \begin{array}{c}
        a \\ b
      \end{array} \right](z | \tau) =
    \sum_{n\in\mathbb Z} e^{i\pi\tau
      (n+a)^2}e^{2i (n+a)(z+\pi b)}
  \end{equation}
}:
\begin{equation}
  \label{eq:basisf}
  \chi_{s+1}^{(0)} = \Psi_{0 s} \equiv \left(\frac{2ql_2}{l_1}\right)^{1/4}e^{fz(z-\overline
    z)}\thetachar{s/q}{0}\left(\frac{2q\pi z}{l_1}
    |iq\frac{l_2}{l_1} \right)
\end{equation}
where $s=0,\ldots , q-1$. They  satisfy the following orthonormality
conditions
\begin{equation}
  \frac{1}{\mathcal A}\int_{\mathbb T^2} d^2 x \; \bar{\chi}^{(0)}_i(x)
  \chi^{(0)}_j(x) = \delta_{i j} 
\end{equation}
Using the operator $(\mathcal T_{v})^{-1}$ one
obtains an associated  basis of the space of solutions Eq.~\ref{BOGONE} 
for any $v$ within one patch of the dual torus. A general solution of
Eq.~\ref{BOGONE} is then given by 
\begin{equation}
  \chi^{(v)} = \sum_{i=1}^{q} c_i \chi^{(v)}_i
\end{equation}
where  $c_i$ are $|q|$ complex constants.  
In principle, one might be led to conclude that 
the  $c_i$  and its complex conjugates,  together with the coordinates
$(v,\overline v)$ provide  coordinates of the moduli space of solutions. 
However, notice that the Higgs and gauge field are invariant
(up to a global gauge transformation) under the replacement:
\begin{equation}
  c_i \longrightarrow \kappa c_i; \qquad h \longrightarrow
  h+log(|\kappa|) 
\end{equation}
where $\kappa$ is an arbitrary complex number. Hence, for fixed $v$
the space of solutions becomes topologically equivalent to the complex projective
space $\mathbb{CP}^{q-1}$ and the coefficients $c_i$ can be regarded as
homogeneous coordinates in this manifold. This completes the structure
of the moduli space as a fibre bundle with fibre $\mathcal F =
\mathbb{CP}^{q-1}$. This 
result is a particular case of the results of Ref.~\cite{Shah:1994us} where
this structure is proved  for vortices in more general
manifolds. This parametrisation gives us the first set of coordinates 
that we will use  
\begin{equation}
  \label{eq:basis1}
  \mathcal B_1 = \{v,\overline v, c_i,\overline c_i\}\quad (i=1,\dots,q)
\end{equation}
A standard  choice to obtain inhomogeneous coordinates is to divide all
the $c_i$ by one of them (say $c_1$). This gives good coordinates in
the patches where $c_1\neq 0$. For the $q=2$ case, since
$\mathbb{CP}^1 \approx S^2$, we will also make use of the standard 
spherical coordinates.

Another quite  natural choice of coordinates of the moduli space 
 is given by the position of the zeros of the Higgs field
$\{\tilde z_r\}$. In Ref.~\cite{Gonzalez-Arroyo:2004xu}, we verified that the
function $\chi^{(0)}$ must have $|q|$ zeros (counted with multiplicity), and
that the centre of mass of these zeros is located in the point
$Z_c^{(0)} = \frac{1}{4}(l_1+il_2)$. Since the application of
the operator $\mathcal T_v$ does not change the number of zeros, the
function $\chi^{(v)}$ will  also possess $|q|$ zeros obeying 
\begin{equation}
  \frac{1}{q}\sum_r \tilde z_r = Z_c^{(0)} + \imath \frac{\overline v}{2f}
\end{equation}
Hence,  the coordinate $v$ of the moduli space can be seen to be
directly given by the  location of the centre of mass of the zeros of the
Higgs field (note that the coordinate $v$ must lie in its proper
range. This can always be achieved using the fact that the position of
the zeros are defined up to  translations by one period). Then, 
one can take as coordinates of the fibre $\mathcal F$, 
the position of the zeros relative to the centre of mass, 
$\omega_i = \tilde z_i - Z_c$  (where $Z_c = \frac{1}{q}\sum_i \tilde z_i$). 
With an appropriate ordering of these relative coordinates,  
we can use the set
\begin{equation}
  \label{eq:basis2}
  \mathcal B_2 = \{v,\overline v, 
  \omega_i, \overline \omega_i
  \}\quad(i=1,q-1)
\end{equation}
as coordinates of the moduli space. 

From this second construction, it is clear that this moduli space is a 
complex manifold having $\overline v$ and $\omega_i$ as possible
complex coordinates. Instead, one might use $\overline v$ and $c_i$. 
The corresponding change of coordinates (see
Ref.~\cite{Gonzalez-Arroyo:2004xu} for  details) is holomorphic.

\subsection{Bradlow parameter expansion}

The main difficulty in obtaining the explicit form of the solutions 
is the non-linear character  of the vortex equation Eq.~\ref{BOGTWO}.
In our previous paper we gave a recipe for solving it  by an expansion 
in the parameter $\epsilon$, directly related to the Bradlow parameter. 
For $\epsilon=0$ the
solution is given by a constant value of $h$. This  constant is fixed
uniquely by the flux condition and the normalisation of $\chi$. In
particular, it vanishes if $\chi$ is chosen of unit norm. In this limit 
the Higgs field vanishes and the magnetic field is uniform.
For small positive values of $\epsilon$, one expects a solution which 
departs only  slightly from a constant. In general, one can write the solution 
of the vortex equation as a power series in $\epsilon$:
\begin{equation}
h=\sum_{n=1}^{\infty} h^{(n)}\epsilon^n 
\end{equation}
Plugging this into Eq.~\ref{BOGTWO} and matching equal powers of $\epsilon$
on both sides, one obtains a series of linear non-homogeneous
equations which allow the determination of the  
functions $h^{(n)}$ iteratively. In practice, for the torus case
the solution of the linear equations can be accomplished by 
expanding  $h^{(n)}(x)$ in  Fourier series, whose coefficients 
can be determined by the iterative procedure mentioned  previously.
The main difficulty is that to obtain these coefficients one has to perform a
increasingly large number of convolutions of the previously determined
$h^{(m)}(x)$. Fortunately, the Fourier modes of these functions fall very 
rapidly and it is possible to compute the dominant ones with 10-15
significant digits by performing these convolutions numerically (truncating
the infinite sums down to finite sums). 

In Ref.~\cite{Gonzalez-Arroyo:2004xu} we were 
able to carry the $\epsilon$-expansion up to order 51 with a fairly modest
computational effort. Having those many terms in the expansion allowed
us  to explore the convergence of the series. It is, of course, impossible in this
way to prove convergence of the series. However, even if the series is
asymptotic, it is possible to test what precision that can be attained for a
particular value of $\epsilon$. Here it is important to notice that by
virtue of its definition, the range of $\epsilon$ values that we are
interested in goes from $0$ to $1$. The upper edge of the interval
$\epsilon=1$, corresponds to the infinite volume limit of the torus: the plane.
The results of our  analysis  turned out to be very 
positive. Meaningful results, comparing well in value and precision with that 
of other authors, were obtained for the worst possible case of the plane 
($\epsilon=1$). Our method, in contrast to these other results, gives us the 
solutions for all torus sizes and can be applied irrespective of the value 
of $q$ or the symmetries invoked. Having an  analytical expansion  for  the
solutions has many advantages over purely numerical methods. In
particular, many different theoretical questions involving  vortices
can be attacked in the same way. This paper develops  one of these
applications, the determination of the dynamics of vortices within the
geodesic approximation and our Bradlow parameter expansion. In
concrete, the goal is to compute the metric in the moduli space of
multi-vortices on the torus. This will be explained in the next
subsection.

\subsection{Derivation of the metric}
We will now allow our fields $A_i$ and $\phi$ to depend on time (while
remaining $z$-independent), but  
we will impose that at any time they correspond to a  configuration 
of minimal energy within the corresponding flux sector($2 \pi q$),
i.e. a solution of the Bogomolny equations. 
 Therefore, the motion is indeed within the moduli space of
solutions of the Bogomolny equations. 

To study dynamics we have to compute the kinetic energy, which in   the
$A_0=0$ gauge, takes the form:
\begin{equation}
  T=\frac{1}{2}\int d^2x\, (\dot{A}_i^2+|\dot{\phi}|^2)
\end{equation}
The potential energy is fixed to be $\pi$ times  the total magnetic
flux, and is time-independent. Thus, the dynamics of the system is given by 
this kinetic term. This amounts to  a finite dimensional classical system, 
which corresponds to geodesic motion within the $2q$-dimensional manifold
of solutions of the Bogomolny equations. To determine the metric, we simply have to 
parametrise $A_i$ and $\phi$ and their time derivatives in terms of whatever 
coordinates of this manifold we choose, and plug them into the expression
for $T$. 

We will be using  the parametrisation of the Higgs and gauge fields 
introduced  previously. In addition to the Bogomolny equations, 
we also need the dynamical equations involving time derivatives. Some can be
deduced by differentiating the Bogomolny equations~\ref{BOGBOTH} with
respect to time. We have 
\begin{subequations}
   \begin{equation}
     (\overline\partial +f z)\dot\chi^{(v)} = \imath (\bar{v}
     \dot{\chi}^{(v)}+ \dot{\bar{v}} \chi^{(v)})
   \end{equation}
   \begin{equation}
     \partial\overline\partial \dot{h} = 
      \dot{h} |\phi|^2- 
     \frac{\epsilon}{2} e^{-2h}\frac{\partial}{\partial t}|\chi^{(v)}|^2
   \end{equation}
\end{subequations}
One has then to add the Gauss constraint (one of Maxwell equations for
the system)
\begin{equation}
  \partial_i\dot{A}_i=\mathop{\rm Im} (\dot{\phi}\overline\phi)
\end{equation}
This equation  can be combined with the time derivative of the second
Bogomolny equation to give a complex equation:
\begin{equation}
  \partial\dot{\overline{A}}=-\imath \overline\phi \dot{\phi}
\end{equation}
In summary,  the only equations that remain to be solved 
are:
\begin{subequations}
  \label{eq:basics}
  \begin{equation}
    \label{eq:basic1}
    \partial\overline\partial h =
    \frac{\epsilon}{2}\left( 1-e^{-2h}\overline\chi^{(v)}\chi^{(v)} \right)
  \end{equation}
  \begin{equation}
    \label{eq:basic2}
    \hat O \dot H = \epsilon e^{-2h}\overline \chi^{(v)}\dot\chi^{(v)}
  \end{equation}
\end{subequations}
Where $\hat O \equiv -\partial\overline\partial + |\phi|^2$. These
equations determine both $h$ and $\dot H$ in terms of $\chi^{(v)}$ and 
$\dot\chi^{(v)}$. Integrating both sides of Eqs.~\ref{eq:basics},
we obtain
\begin{subequations}
  \label{eq:basicscte}
  \begin{equation}
    \frac{1}{\mathcal
      A}\int d^2x\,e^{-2h}\overline{\chi^{(v)}}\chi^{(v)} = 1
  \end{equation}
  \begin{equation}
    \int d^2x\,e^{-2h}\overline{\chi^{(v)}}\dot\chi^{(v)} = 
    \int d^2x\,e^{-2h}\overline \chi^{(v)} \dot{H}\chi^{(v)}
  \end{equation}
\end{subequations}
that are basic equations for the constant pieces of $h$ and $\dot
H$. Eqs.~\ref{eq:basics} and \ref{eq:basicscte}  uniquely determine
$h$ and $\dot H$ given $\chi^{(v)}$ and $\dot\chi^{(v)}$. Hence, 
parametrising  $\chi^{(v)}$ in terms of certain  coordinates of the
moduli space, one can then  use  Eqs.~\ref{eq:basics},
\ref{eq:basicscte}, to express  
$h$ and $\dot H$ in terms of these  coordinates and their time derivatives.

Using our parametrisation, we might express the kinetic term 
as 
\begin{eqnarray}
  \label{eq:Texpr}
    T  & = &  \frac{1}{2} \int d^2x\, \left\{
      |\dot v|^2 + \epsilon e^{-2h}\left| \dot\chi^{(v)} \right|^2
      - \dot{\overline H} 
       \hat O\dot H
    \right\}\\
    \nonumber
    & = & \frac{1}{2} \int d^2x\, \left\{
      |\dot v|^2 + \epsilon e^{-2h} \dot{\overline{\chi}}^{(v)} 
      \left( \dot{\chi}^{(v)} - \dot{H}_v \chi^{(v)}\right)
    \right\}
\end{eqnarray}
This is the crucial formula  from which the metric and its
perturbative expansion can be obtained. It is not hard to show that $T$  
 is invariant under the replacement of $\dot\chi^{(v)}$ by
$\dot\chi^{(v)} - \lambda \chi^{(v)}$ for any constant $\lambda$. 
This expresses the irrelevance of the normalisation of $\chi^{(v)}$.  
In Eq.~\ref{eq:Texpr} the time derivatives of the coordinates appear
in  $\dot{v}$ and $\dot\chi^{(v)}$. These two terms  are not
independent variations within our fibre bundle choice of
coordinates. One should rather use $\dot{v}$ and $\dot\chi^{(0)}$. For
that purpose we need the formula 
\begin{equation}
\label{eq:chiv}
  \dot\chi^{(v)} = \frac{d}{dt}\left(
    \left(\mathcal T_v\right)^{-1}[\chi^{(0)}]
  \right) = (\mathcal T_v)^{-1}[ \dot\chi^{(0)} 
  -\frac{\imath \dot{\overline v}} {2f}(\partial -f\overline z + iv)
  \chi^{(0)} ] 
\end{equation}
Using this expression we might also relate the solution of
Eq.~\ref{eq:basic2} for $v\ne 0$, which we will note $\dot{H}_v$, with
the corresponding one for $v=0$:
\begin{equation}
  \label{eq:dotHv}
  \dot{H}_v\left(z +\frac{i \overline v}{2 f}, \overline z -\frac{ i v}{2 f}\right)=
  \dot{H}\left(z,\overline z\right) - \frac{i \dot{\overline{v}}}{2 f}
  \left(\delta(z,\overline z) +iv\right)
\end{equation}
where $\delta$ is the solution of the equation
\begin{equation}
   \hat O \delta = \epsilon e^{-2h}\overline \chi^{(0)}(\partial
   -f\overline z )\chi^{(0)}
\end{equation}
Now we should plug these relations  into the formula for the kinetic
term $T$, and after changing variables in the integrals and decomposing 
$\chi^{(0)}$ in the aforementioned basis,  we arrive 
at an expression of the form
\begin{equation}
  T = \frac{\pi}{2}\left(g_{vv}(c_i)\dot v\dot {\overline v} + 
         g_{iv}(c_i)\dot {\overline c_i}\dot {\overline v} + 
         g_{vi}(c_i)\dot v\dot  c_i + 
         g_{ij}(c_i)\dot c_i\dot {\overline c_j}\right)
\end{equation}
Where $g_{\alpha\beta}$ is the induced
metric in the moduli space (we follow the normalisation of 
Ref.~\cite{Manton:1993tt}). Furthermore, $g_{iv}$ and $g_{v i}$ vanish.
To show this we focus on  the term proportional to $\dot{v}$ in the 
expression of $T$. It has the form
\begin{equation}
\frac{ i \dot{v}}{4 f} \int d^2x \, I
\end{equation}
where
\begin{equation}
  I=(\overline \partial - f z)\overline \chi^{(0)} \left[\dot
    \chi^{(0)} - \chi^{(0)} \dot{H} \right] \epsilon e^{-2h} = 
  -\overline \partial \partial \overline \partial \dot{H} +\epsilon
  \overline \partial \dot{H} -2 \partial (\overline \partial \dot{H}\,
  \overline \partial h)  
\end{equation}
This, being the derivative of a periodic function, has a vanishing
integral. By hermiticity the same holds for the term proportional to
$\dot {\overline v}$.  In a similar fashion (by partial integration),  
one can show that the term proportional to $|\dot{v}|^2$ is a constant,
to be given below.  Finally, we arrive to the following expression 
for the metric
\begin{equation}
\label{eq:metric1}
  ds^2 = \frac{{\mathcal A}^2 }{4 \pi^2 q}\, dvd\overline v +
  \frac{\epsilon}{\pi}\int d^2x\, e^{-2h}\left\{ 
    |\dot\chi^{(0)}|^2 - \epsilon \dot{\overline\chi}^{(0)}
    \chi^{(0)}\left(\hat O^{-1}\right)  
    \left( e^{-2h}{\overline\chi}^{(0)}
      \dot\chi^{(0)}\right)\right\}
\end{equation}
The first term is the metric associated to the dual torus, while the
second is the metric of the fibre $ \mathcal F$. This factorised
structure was previously found in Ref.~\cite{Shah:1994us}.

From now on we might focus upon the metric of the fibre $\mathcal F$, 
which can be written as 
\begin{equation}
\label{eq:metric2}
 ds^2 = \frac{1}{\pi} \int d^2 x\, \epsilon e^{-2h}
 \dot{\overline{\chi}}^{(0)}
     \left( \dot{\chi}^{(0)} - \dot{H} \chi^{(0)}\right) =
     \frac{-1}{\pi} \int d^2 x\, \partial \left(
     \frac{\dot{\overline{\chi}}^{(0)}}{\overline{\chi}^{(0)}}
     \overline \partial \dot{H} \right)		     
\end{equation}
Notice that, this time, the integral of a partial derivative does not
vanish, because the quantity inside parenthesis has singularities (poles) 
at the zeroes of $\overline{\chi}^{(0)}$. Hence, the metric is
expressed by the value of this function in the vicinity of these
zeroes. This property was discovered in Ref.~\cite{Samols:1991ne}
for vortices on the plane. In particular, the metric is twice the 
sum of the residues of the quantity inside brackets of the last equality
at its poles.

Our method of computation of the metric follows from
 substituting into Eq.~\ref{eq:metric2} the expression of
  $h$ and $\dot H$ as a power expansion in powers of $\epsilon$. 
This expansion follows from solving Eqs.~\ref{eq:basics} and
\ref{eq:basicscte} order by order in $\epsilon$. 
The resulting expression of the metric in the fibre ${\mathcal F}$ 
has the form 
\begin{equation}
  g_{ij}(c_k)= \sum_{N=0}^\infty g_{ij}^{(N)}(c,\overline c)\epsilon^N
\end{equation}
The computation of the first two orders is detailed in 
Appendix~\ref{ap:order2}. The first non-zero term in this
series, is given by
\begin{equation}
  g_{ij}^{(1)} = \frac{\mathcal A}{\pi||c||^2}\left(
    \delta_{ij} - \frac{c_i\overline c_j}{||c||^2}
  \right)
\end{equation}
where $||c||^2 = \sum_i |c_i|^2$. This is just a multiple of the 
Fubini-Study metric. This result  agrees with  the metric obtained 
in this limit in Ref.~\cite{Baptista:2002kb}, 
but for vortices living in the two sphere $S^2$. Using this result 
we can compute the volume of the fibre.
\begin{equation}
\mbox{\rm Vol}({\mathcal F})=  \frac{ (\epsilon {\mathcal
A})^{q-1}}{(q-1)!}
\end{equation}
This value times the volume of the base-space dual torus matches the 
calculation done in Ref.~\cite{Manton:1993tt}, and generalised in 
Ref.~\cite{Manton:1998kq} to other manifolds. Furthermore, the result 
is valid to all orders in $\epsilon$.

In  Appendix~\ref{ap:order2}  the second order is explicitly computed. 
The result involves certain constants which can be expressed as 
an  infinite sum over two integers, or as integrals of theta functions 
with characteristics. The general structure of the metric to all  
orders will be  described in the next subsection. It includes certain  
 constants given by  an increasing number of infinite 
sums over integers. Fortunately, these sums are very fastly convergent
and a few  terms suffice to obtain a double-precision 
($\approx 16$ decimal places) numerical determination.
This fact enabled  our study of the $q=2$ case up to fortieth order,
to be presented in Section~\ref{sc:plane}.

\subsection{Symmetries and general properties of the metric.}

Here we will focus upon the  metric on the fibre 
$\mathcal{F}$ expressed in homogeneous coordinates $c_i$. 
Some of its  properties, which were 
known before, hold order by order in our expansion. For example, 
we know that the metric is K\"ahler~\cite{Samols:1991ne}. As mentioned
previously, we also know  the total volume of the
fibre~\cite{Manton:1998kq}. Since the volume is already obtained at leading
order, this tells us that the higher order contributions do not
contribute to it. 

Another property of the functions $g_{ i j}(c,\overline c)$ has to do 
with the homogeneous nature of the coordinates, and has been mentioned 
previously. Translating the property to the expansion coefficients 
we have:
\begin{equation}
  \label{eq:transv}
  \sum_j c_j g^{(N)}_{ i j} =0
\end{equation}

A more specific    property of the terms $g_{ i j}^{(N)}$  follows from 
the $\epsilon$-expansion of Eqs.~\ref{eq:basics} and
\ref{eq:basicscte}: Each new order in $\epsilon$ is accompanied in the 
right-hand side by a power of $\overline{c}_i c_j/||c||^2$. Analysing
in detail the different equations, we conclude that 
$||c||^2 g_{ i j}^{(N)}$  is a polynomial of degree 2N in the variables 
$\frac{c_i}{||c||}$ and its conjugates. This property is quite crucial 
for our expansion, since it allows the exact determination of the 
metric to any finite order in terms of a finite (and small for $N$
small) number of coefficients. Combining this fact with the previous
properties, a reduction in the number of free coefficients follows. For
example, to leading order we have a polynomial of  degree 2 in
$\frac{c_i}{||c||}$. Property Eq.~\ref{eq:transv}, reduces the
indetermination to  a single overall pre-factor, which is fixed by the 
 volume. Similar simplifications occur for the 
second order result, described in the Appendix.
 
A further reduction in the number of free coefficients follows from 
symmetries. These are  isometries, which have an origin in a geometrical
symmetry of the problem: transformations that leave 
the torus invariant. This includes reflections and rotations by $\pi$. 
If the shape  of the torus is square $l_1=l_2$, we also have rotations by 
$\frac{\pi}{2}$. In addition, we have translations. This might look
surprising, since we are restricting to a fibre which has a fixed
position of the centre of mass of the vortices. However, 
translations by a fraction $1/q$ of the total length in either  direction
do not change the position of the centre of mass. This is a finite 
group with $q^2$ elements. 

In order to exploit the consequences of the symmetries one has to make 
a choice of the basis $\chi^{(0)}_i$. Here, we will give the results
in the basis $\chi^{(0)}_{s+1}=\Psi_{0 s}$ (defined in Eq.~\ref{eq:basisf}). 
The explicit properties of these functions 
needed to express the requirements of the symmetries are  explained 
in the Appendix of Ref.~\cite{Gonzalez-Arroyo:2004xu}. Some subtlety is
necessary. This has to do with the fact that transition functions are, 
in some cases, not invariant under the  Euclidean transformations that 
are candidate symmetries. To recover the symmetry one has to 
accompany the  geometrical transformation  with  a gauge transformation, designed to leave
the transition functions invariant.  We will now present the results. 
The proofs are simple (after consulting
Ref.~\cite{Gonzalez-Arroyo:2004xu}), and are left to the reader.

The invariance under translations along the x axis, implies  that the 
metric is invariant under the replacement 
\begin{equation}
c_s \longrightarrow e^{-2 \pi i s/q} c_s
\end{equation}
Translations along the y axis are associated with the transformation
\begin{equation}
  c_s \longrightarrow c_{s+1}
\end{equation}
where $c_{q}=c_0$. Reflections with respect to the $x_1=0$ line are
associated with the transformation
\begin{equation}
  c_s \longrightarrow \overline c_s
\end{equation}
Finally the $\pi/2$ rotations are associated with the transformations:
\begin{equation}
  c_s \longrightarrow \sum_{s'=0}^{q-1}  \frac{1}{\sqrt{q}} e^{2\pi i s
  s'/q} c_{s'}
\end{equation}
The square of this transformation provides the rotation by $180$ degrees.
In the next section we will make use of these symmetries for the $q=2$
case.

\section{Two-vortex dynamics}
\label{sc:results}

Here we will study with higher detail the specially interesting case
of two vortex dynamics. As mentioned previously, the fibre ${\mathcal F}$
of the moduli space is diffeomorphic to the two-sphere. In the first 
part of this section we will explain  the different coordinates that 
we will be using. Then we will present our results focusing on two
different regimes. For volumes not too far from the Bradlow limit, a 
few orders in the expansion  provide a good description of the metric
and dynamics. It turns out that the metric is fully determined by
fixing a small number of real constants. After explaining this general
structure, we give the numerical value of the 17 constants necessary 
to fix the metric up to fifth order. With this information  the 
corresponding two-vortex dynamics is analysed. On the opposite extreme
we have the case of very large torus sizes.
In this regime, we expect our results to match 
with the case of vortices in the plane. By analysing the results up to
40th order in the expansion  for a $40\times40$  grid of points in the moduli space 
 ${\mathcal F}$, we verify that this is the case. Our method proves 
 to be strongly competitive in extracting information about the metric and
 scattering  of vortices on the plane.

Our formalism,  given in the previous section, expressed the metric in 
homogeneous coordinates:
\begin{equation}
  ds^2 = g_{ij} d\overline c_i dc_j
\end{equation}
and deduced certain properties for the hermitian matrix $g_{ij}$, and
its expansion in powers of $\epsilon$.  In particular, 
the metric function satisfies $c_1 g_{i 1}+ c_2 g_{i 2}= 0$.
Going over to  the non-homogeneous complex coordinate $Z=\frac{c_2}{c_1}$,
we get
\begin{equation}
  \label{eq:deff}
  ds^2 = \frac{\tilde f(Z,\overline
      Z)}{\left(1+|Z|^2\right)^2}d\overline Z d Z
\end{equation}
where  $\tilde f(Z,\overline Z) = ||c||^2\text{Tr}(g) =
||c||^2(g_{11}+g_{22})$. It is also useful to express the metric in 
spherical coordinates $(\theta,\varphi)$, defined by  the following
change of variables:
\begin{subequations}
  \label{eq:basistp}
    \begin{equation}
        \frac{2a}{1+|Z|^2} = \sin\theta\cos\varphi
	  \end{equation}
	    \begin{equation}
	        \frac{-2b}{1+|Z|^2} = \cos\theta
		  \end{equation}
		  \end{subequations}
where $a$, and $b$ are the real and imaginary parts of $Z$ ($Z =
a+\imath b$). In these coordinates the metric takes the form:
\begin{equation}
  \label{eq:basistpm}
  ds^2 = \frac{\tilde f_s(\theta,\psi)}{4}\left(d\theta^2 + \sin^2\theta\, d\varphi^2\right)
\end{equation}
where,  the function  $\tilde f_s$ is the same as $\tilde f$ but
expressed in the new variables. 

We will also use the more conventional coordinates of the moduli space: 
the zeros of the Higgs field. Let $z=u$ be the location of one of the zeroes 
(the position of the other zero is given by the centre of
mass condition). In order for the change of variables to remain
$\epsilon$-independent, we will consider first $w$, the  position of the zero relative
to the torus, given by:
\begin{equation}
  w = \frac{4\pi }{l_1} u
\end{equation}
Now, we can make use of the expressions given in Ref.~\cite{Gonzalez-Arroyo:2004xu}
to give the holomorphic change of coordinates:
\begin{equation}
  Z(w) = -\frac{\vartheta_3(w|\imath 2\tau)}{\vartheta_2(w|\imath
  2\tau)}= -\frac{\Psi_{0 0}(w)}{\Psi_{0 1}(w)}
\end{equation}
where $\vartheta_i(z|t)$ are the $i^{th}$ classical Jacobi theta
functions\footnote{All of the four classical Jacobi theta functions
  can be defined in terms of the theta function with characteristics
  (see footnote~\ref{ft:theta}, page~\pageref{ft:theta})
  \begin{subequations}
    \begin{equation}
      \vartheta_3(z|t) = \thetachar{0}{0}(z|t)
    \end{equation}
    \begin{equation}
      \vartheta_1(z|t) = -\thetachar{1/2}{1/2}(z|t)
    \end{equation}
    \begin{equation}
      \vartheta_2(z|t) = \thetachar{1/2}{0}(z|t)
    \end{equation}
    \begin{equation}
      \vartheta_4(z|t) = \thetachar{0}{1/2}(z|t)
    \end{equation}
  \end{subequations}
  See Ref.~\cite{ww:analysis} for more detailed information about  Jacobi theta functions.
} 
and $\tau=l_2/l_1$. For ease of notation in the next formulas we 
will omit  the second argument of  theta functions,  assuming  
it is equal to  $\imath 2\tau$). 
Finally, we arrive  to the form of the metric in the  relative position
coordinate $w$: 
\begin{equation}
\label{eq:metricw}
  ds^2 = \tilde f_w(w,\overline w)|\vartheta_4(0)|^4\frac{|\vartheta_1(w)|^2
    |\vartheta_4(w)|^2}{\left( |\vartheta_2(w)|^2
      + |\vartheta_3(w)|^2 \right)^2}\, d\overline wdw
\end{equation}
where $\tilde f_w$ is obtained from $\tilde f$ by change of variables.
From here, one can trivially re-express the metric using $u$ and
$\overline u $ as coordinates.

After the presentation of the coordinates we will now  deduce  the 
implications for the function $\tilde f$ of the  properties
of the metric studied in the previous section (The K\"ahler property 
is obvious in our case). Let us examine how  the symmetry transformations 
act on the variable $Z$. Translations along both  axis
lead to the transformations $Z \longrightarrow -Z$ and
 $Z \longrightarrow 1/Z$.  Reflection invariance leads  
 to the transformation property  $Z \longrightarrow \overline Z$.
 Hence, the function $\tilde f(Z,\overline Z)$ must be
  even and symmetric under the exchange of its arguments and satisfy
  \begin{equation}
  \tilde f(Z,\overline Z) =\tilde f\left(\frac{1}{Z},\frac{1}{\overline Z}\right)
  \end{equation}
Finally, the transformation associated with $\pi/2$ rotations (if
$l_1=l_2$) corresponds to 
\begin{equation}
Z \longrightarrow \frac{1-Z}{1+Z}
\end{equation}
implying that $\tilde f$ must be invariant under this change of
variables.

\subsection{Near the Bradlow limit}
\label{sc:bradlow}

Here we will examine the   first few orders  of the expansion of 
$\tilde f$ in powers of $\epsilon$. These terms will describe 
quite accurately the structure 
of the metric for small values of $\epsilon$ or, equivalently, for
torus sizes which are not too large compared with the critical area of 
Bradlow. 

As explained in the previous section, the 
contribution  of order $\epsilon^N$ to the metric tensor 
$||c||^2 g^{(N)}_{i j}$ is a polynomial of degree $2N$ in $c_i/||c||$.
Going over to the variable $Z$, we conclude that the Nth order contribution
to $\tilde f(Z,\overline Z)$ is given by 
\begin{equation}
  \tilde f^{(N)}(Z,\overline Z) =\frac{P_N(Z,\overline Z)}{(1+|z|^2)^{(N-1)}}
\end{equation}
where $P_N(Z,\overline Z)$ is a  polynomial of degree $2N-2$ in its
arguments. The symmetries imply certain restrictions on the
coefficients of this polynomial. Writing $Z=a+ib$, we conclude that
$P_N(Z,\overline Z)$ is a real polynomial in $a^2$ and $b^2$. This brings 
down the number of coefficients to $N(N+1)/2$. Further restrictions 
follow from the symmetry $Z \longrightarrow 1/Z$. It is easier to
express them in terms of the variables $\theta$ and $\varphi$ defined 
in Eq.~\ref{eq:basistp}. One  
concludes that $\tilde f_s^{(N)}(\theta, \varphi)$ is a polynomial in 
$\cos^2(\theta)$ and $\sin^2(\theta)\sin^2(\varphi)$ of degree $(N-1)$.
This is exemplified by the first two orders, explicitly calculated   in
Appendix~\ref{ap:order2}.  The leading order term reduces in our case to the 
a constant curvature metric on the sphere of radius 
$\sqrt{\frac{\epsilon \mathcal A}{4\pi}}$. 
The next order is given  by a sum of three pieces, which  are related 
by the condition that the total contribution to the area is zero. 

We will now restrict to the most symmetric case $l_1=l_2$. 
The additional symmetry resulting
from 90 degree rotations, implies  the following general form:
\begin{equation}
  \label{eq:expansi}
      \tilde f_s^{(N)}(\theta,\varphi) = \frac{\mathcal{A}}{\pi}\, \sum_{j, k=
          0}^{j+2k \le N-1} A^{(N)}_{jk} \cos^{2j}{\theta}
	      \sin^{4k}{\theta} \cos^{2k}{2\varphi}
	      \end{equation}
where $A^{(N)}_{jk}$ are real coefficients.	      
It is relatively easy to compute these 
coefficients numerically up to machine double precision. 
The results up to fifth order  of the expansion are collected
in table~\ref{tab:metric}.

\begin{table}
  \centering
  \begin{tabular}{|c|c||c|c|}
    \hline
    & $\mathbf{O(3)}$ & & $\mathbf{O(2)}$ \\
    \hline
    $A^{(3)}_{00}$ & $1.57\ 24\ 57\ 43\ 63\ 76\ 56(1)\times 10^{-1}$ &
    $A^{(2)}_{00}$ & $2.09\ 85\ 65\ 63\ 87\ 72\ 34\ 78\ 07\times 10^{-1}$ \\
    \hline
    $A^{(3)}_{10}$ & $-7.35\ 97\ 58\ 95\ 32\ 71\ 25\times 10^{-1} $ & 
    $A^{(2)}_{10}$ & $-6.29\ 56\ 96\ 91\ 63\ 17\ 04 39\ 78\times 10^{-1}$ \\
    \hline
    $A^{(3)}_{20}$ & $4.40\ 39\ 77\ 74\ 02\ 36\ 00\times 10^{-1}$ & 
    & $\mathbf{O(5)}$ \\
    \hline
    & $\mathbf{O(4)}$ & $A^{(5)}_{00}$ & $3.24\ 86\ 80\ 02\ 33\ 12\ 76\times
    10^{-1}$ \\
    \hline
    $A^{(4)}_{00}$ & $2.56\ 44\ 31\ 24\ 38\ 37\ 31\ 64(1)\times 10^{-1}$&
    $A^{(5)}_{10}$ & $-1.59\ 28\ 19\ 34\ 17\ 36\ 19\ 6$\\
    \hline
    $A^{(4)}_{10}$ & $-1.09\ 63\ 43\ 50\ 68\ 01\ 26(2)$&
    $A^{(5)}_{20}$ & $2.35\ 30\ 76\ 79\ 79\ 16\ 14\ 0$\\
    \hline
    $A^{(4)}_{20}$ & $1.12\ 62\ 17\ 59\ 37\ 90\ 85(1)$&
    $A^{(5)}_{30}$ & $-1.33\ 17\ 87\ 41\ 82\ 85\ 50\ 3$\\
    \hline
    $A^{(4)}_{30}$ & $-3.23\ 47\ 12\ 72\ 58\ 54\ 03(4)\times 10^{-1}$&
    $A^{(5)}_{40}$ & $2.44\ 37\ 72\ 51\ 19\ 98\ 28\times 10^{-1}$\\
    \hline
    $A^{(4)}_{01}$ & $-2.62\ 60\ 73\ 46\ 53\ 83\ 54(1)\times 10^{-1}$&
    $A^{(5)}_{01}$ & $-4.46\ 53\ 65\ 89\ 63\ 78\ 00\times 10^{-1}$\\
    \hline
     & &
    $A^{(5)}_{11}$ & $4.62\ 92\ 29\ 53\ 30\ 50\ 51\times 10^{-1}$\\
    \hline
  \end{tabular}
  \caption{Numerical coefficients which determine the two-vortex
    metric up to fifth order for $l_1=l_2$.}
  \label{tab:metric}
\end{table}

From Eq.~\ref{eq:expansi} one sees that  the number of terms grows 
like $N^2/4$, although from the table we see that some coefficients
turn out to be 0. In addition,  there are  relations among them. 
For example,  $A^{(2)}_{10}=-3A^{(2)}_{00}$. This is a particular 
case of the  restriction following from the vanishing of the  
volume contribution beyond the leading order. This leads to the
following relations among the coefficients:
\begin{equation}
\sum_{ j k } A_{jk}^{(N)}  \frac{\Gamma\left( j+\frac{1}{2} \right)
\left[(2k)!\right]^2}
{(k!)^22^{2k+1}\Gamma\left(j+2k+\frac{3}{2}\right)}=0
\end{equation}
which are nicely satisfied by our coefficients up to machine double
precision.  This provides a non-trivial check of our numerical
determination. Alternatively to Eq.~\ref{eq:expansi} one could have used
an expansion in spherical harmonics. The volume condition implies 
that the coefficient of $Y_{0 0}(\theta, \varphi)$ should vanish.

We will now analyse the two-vortex dynamics that follows from our 
fifth order metric. This is governed by the Hamiltonian
\begin{equation}
  H = \frac{1}{\tilde f(\theta,\varphi)}\left(
    p_\theta^2 + \frac{p_\varphi^2}{\sin^2\theta}
  \right)
\end{equation}
The equations of motion are
\begin{subequations}
  \begin{eqnarray}
    \dot\theta & = & \frac{2 p_\theta}{\tilde f} \\
    \dot\varphi& = &\frac{2 p_\varphi}{\tilde f\sin^2\theta} \\
    \dot p_\theta & = & \frac{1}{\tilde f}\left( 
      E\frac{\partial\tilde f}{\partial\theta} +
      2 p_\varphi^2\frac{\cos\theta}{\sin^3\theta} 
    \right) \\
    \dot p_\varphi & = &\frac{E}{\tilde f}\frac{\partial\tilde
      f}{\partial\varphi}
  \end{eqnarray}
\end{subequations}
Where $E$ is the energy of the system, that is conserved. These 
equations can be easily integrated numerically to obtain the
trajectories in vortex moduli space.  

Notice, that up to third order, the metric is independent of 
$\varphi$. Hence, $p_\varphi$ is 
a conserved quantity, and the problem is  integrable. 
In this case, the equation for the trajectory in the moduli
space is given by
\begin{equation}
  \varphi_f = \varphi_0 +
  \int_{\theta_0}^{\theta_f}\frac{d\theta}{\sin^2{\theta}
    \sqrt{\frac{\tilde f(\theta)}{b^2} - \frac{1}{\sin^2{\theta}}}}
\end{equation}
where $b=\frac{p_\varphi}{\sqrt{E}}$.

\begin{figure}[htbp]
  \centering
  \mbox{
    \subfigure[Poincare map for $\epsilon=0.4$ and $E=60.0$ for the first 3
    orders.]{\includegraphics[width=7.55cm]{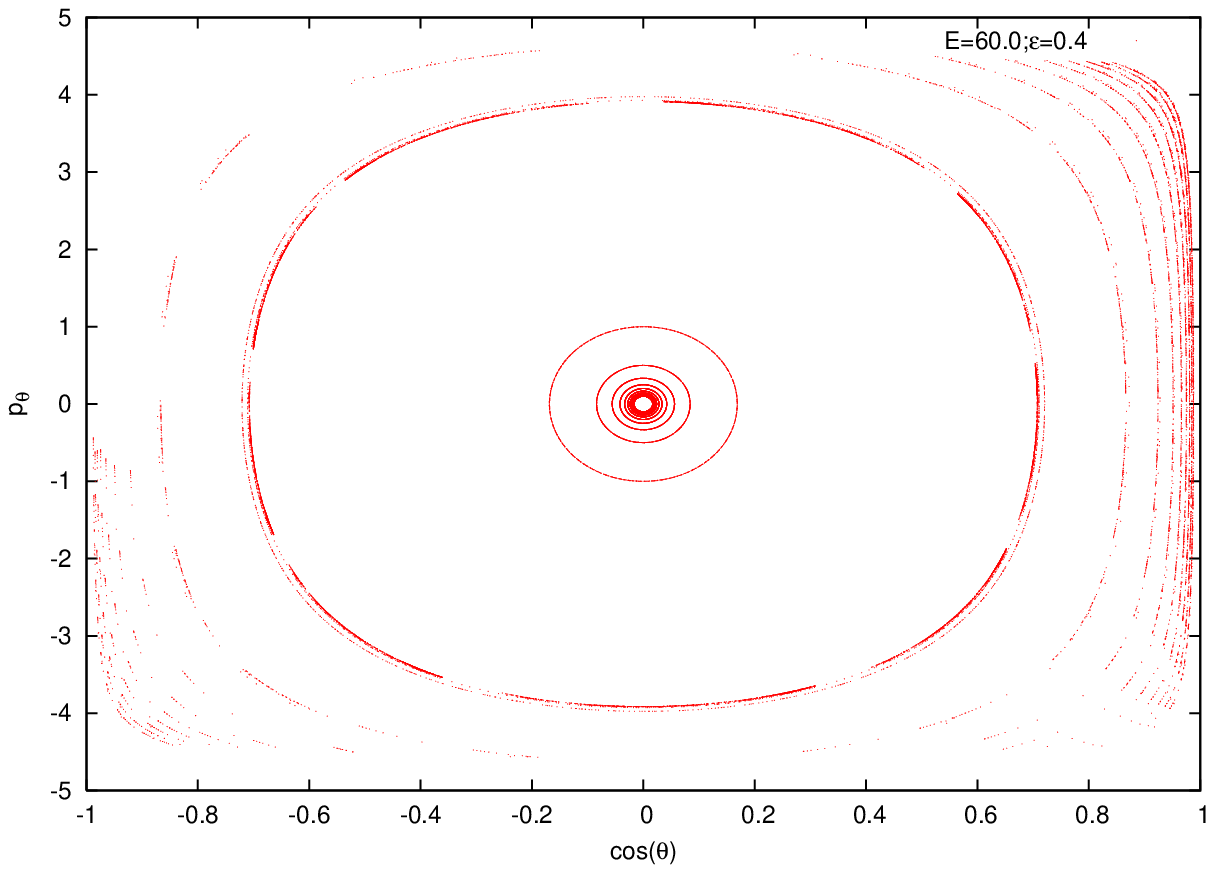}}
    \subfigure[Poincare map for $\epsilon=0.4$ and $E=60.0$ for the first 5
    orders.]{\includegraphics[width=7.55cm]{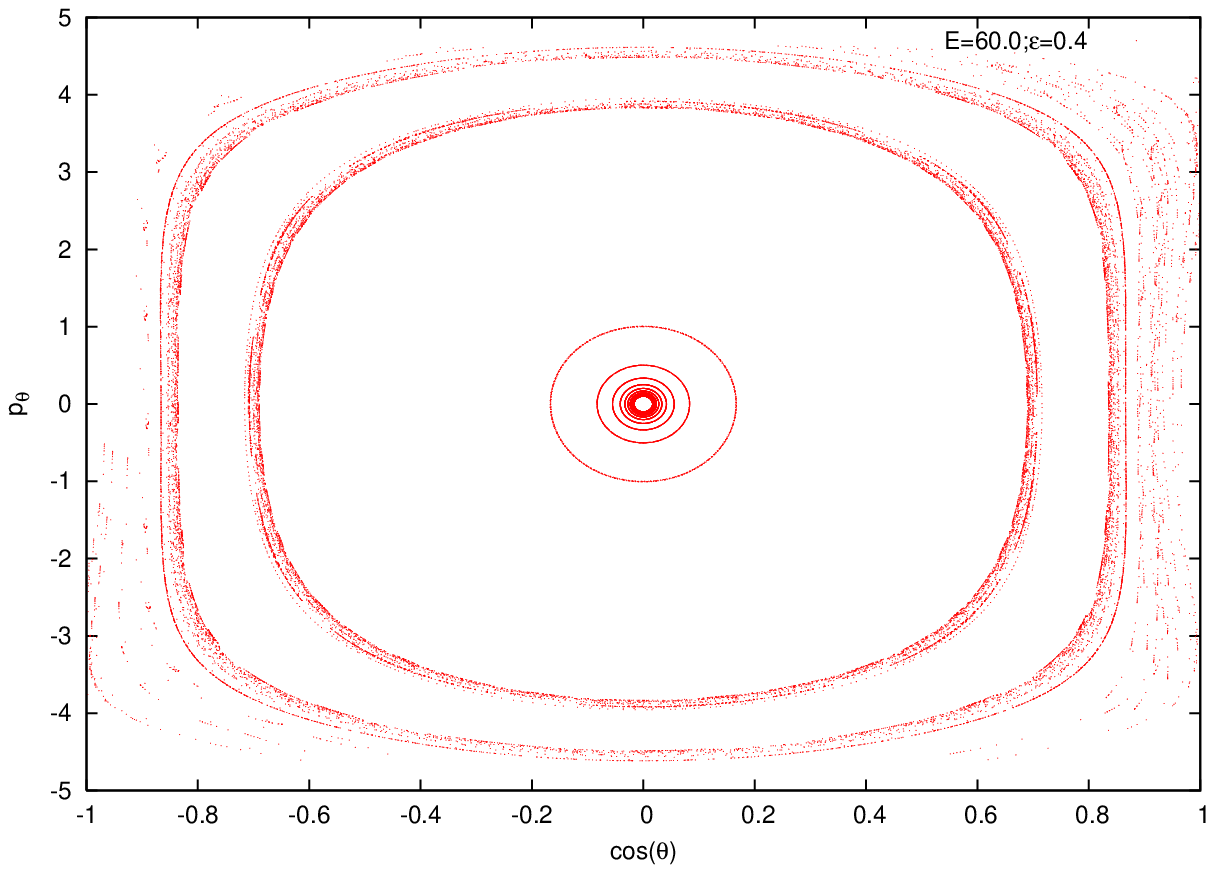}}
  }
  \mbox {
    \subfigure[Detail of the Poincare map for $\epsilon=0.4$ and
    $E=200.0$ for the first 3
    orders.]{\includegraphics[width=7.55cm]{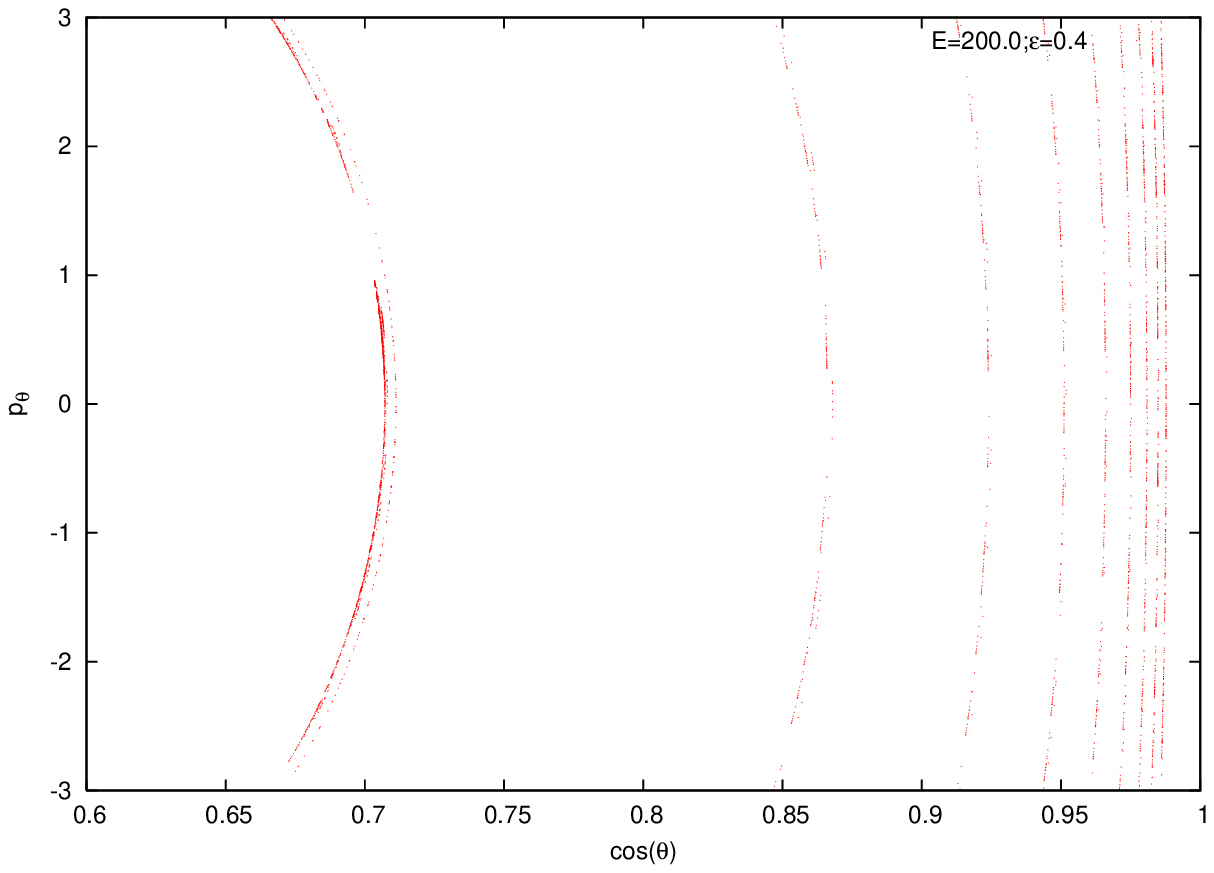}} 
    \subfigure[Detail of the Poincare map for $\epsilon=0.4$ and
    $E=200.0$ for the first 5
    orders.]{\includegraphics[width=7.55cm]{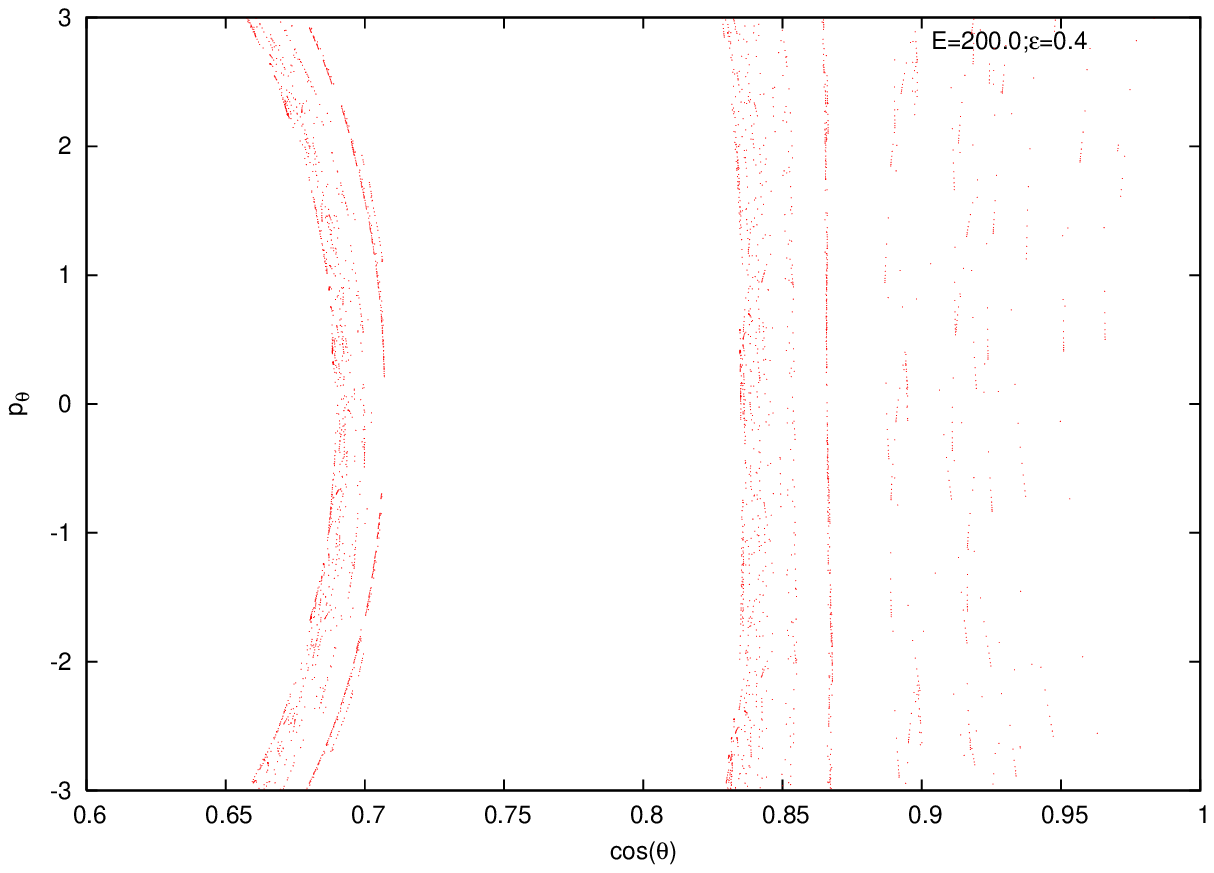}} 
  }
  \caption{Examples of Poincare maps for the dynamics near the Bradlow limit.}
  \label{fig:poincare}
\end{figure}

An interesting mathematical question is whether the two vortex 
dynamics on the torus is integrable for all values of $\epsilon$. 
As we have seen, this is the case up to third order in the expansion.
In the next subsection we will mention that it is also the case in the
$\epsilon \longrightarrow 1$ limit, since then we recover rotational 
invariance and angular momentum becomes a conserved quantity. Our metric
calculations could be used to give a numerical and/or  analytical answer 
to this question. We will not attempt to study that point here. Instead 
we will address the much simpler problem of obtaining and describing 
the Poincare maps obtained from the third order and fifth  order
metric. Results are displayed in Fig.~\ref{fig:poincare} for two values 
of $\epsilon$. We see signs of how some  invariant-tori  seem to be
destroyed, pointing to the non-integrability of the metric truncated 
to fifth order. This is not surprising and does not conflict with the 
conjectured integrability for all values of $\epsilon$. Our results, 
nonetheless, can be considered  as a first step, which might help in 
pointing  to the relevant region of phase space where one should  look.

\subsection{Dynamics of vortices in the plane ($\epsilon\rightarrow 1$
  limit)}
\label{sc:plane}

The dynamics of the Abelian Higgs model in $\mathbb R^2$ is a well
studied problem, both analytically~\cite{Samols:1990kd,Samols:1991ne}
and numerically~\cite{Moriarty:1988fx,Samols:1991ne}. Following the notation 
of Ref.~\cite{Samols:1991ne} we might write
\begin{equation}
  ds^2 = 8 F^2(2|u|)\ dud\overline{u}
\end{equation}
Once the function $F(2 |u|)$ is obtained, the dynamics in the geodesic
approximation is completely determined. The form of the function
$F(2|u|)$  has been deduced analytically for the special case of
asymptotically separated vortices~\cite{Manton:2002wb}, and has also
been studied numerically several
times\cite{Samols:1990kd,Samols:1991ne}. In this paper   we will 
use our expansion method  to obtain a precise determination 
of $F(2 |u|)$ for all values of $|u|$. 

We will naturally assume that, if the torus is large enough
compared both with the typical size  of  vortices, and with the
distance between them, the \emph{dynamics} will be accurately described 
by the one in $\mathbb R^2$. This means that, at least formally,
we can calculate the desired function $F(2|u|)$ by taking the $\epsilon
\longrightarrow 1$ limit while keeping $u$ fixed. At the beginning of
this section we showed how to express the metric in the relative torus
position coordinate $w$. The formula (Eq.~\ref{eq:metricw}) involves 
$\tilde f_w$ times a known function arising from the  change of variables. 
We will first study the behaviour of the latter  factor in the  
appropriate limit ($\epsilon \longrightarrow 1\, , \ u\  \text{fixed}$).
Restricting to the $l_1=l_2=l$ case and recalling that $w=u \sqrt{2
\pi (1-\epsilon)}$ we get:
\begin{equation}
  \left( \frac{4 \pi }{l} \right)^2
    |\vartheta_4(0)|^4\frac{|\vartheta_1(w)|^2
        |\vartheta_4(w)|^2}{\left( |\vartheta_2(w)|^2
	      + |\vartheta_3(w)|^2 \right)^2}
	        \stackrel{\epsilon \rightarrow 1}{\longrightarrow}
		  \frac{16 \pi^2Z_2^2}{(1+Z_0^2)^2}(1-\epsilon)^2|u|^2
		  \end{equation}
		  where\footnote{For the rectangular symmetric case
		  $l_1=l_2$, the
		    numerical values of $Z_0$ and $Z_2$ are
		      \begin{eqnarray*}
		          Z_0 & = & -1-\sqrt{2} \\
			      Z_2 & = & -1.18\ 91\ 73\ 93\ 79\ 71\ 08\ 15
			        \end{eqnarray*}
				} $Z_2=-\frac{1}{2}\frac{\vartheta_3(0)\vartheta_4^4(0)}
				{\vartheta_2(0)}$, and $Z_0 = -\vartheta_3(0)/\vartheta_2(0)$.

In summary, the function $F^2(2|u|)$ can be recovered as 
\begin{equation}
\label{eq:Flimit}
F^2(2|u|) = \lim_{\epsilon \longrightarrow 1}\ (1-\epsilon)^2 \tilde
f_w(u \sqrt{2 \pi (1-\epsilon)}, \overline u \sqrt{2 \pi (1-\epsilon)}, \epsilon)\ 
 \frac{2\pi^2Z_2^2}{(1+Z_0^2)^2} |u|^2 
\end{equation}
 In practice, to obtain an approximation to 
$F^2(2|u|)$ from our expansion we proceeded as follows. We computed the
coefficients in the expansion in $\epsilon$ of the metric up to 40th
order, for a regular rectangular array of points in the complex plane of the 
variable $w$. The range of values is
that of a complex torus. Due to symmetries (which we checked) it is enough
to take values in one quadrant. We considered a grid of $40 \times 40$
points. Getting to order $40$ is feasible  by implementing the
iterative solution  of the equations  on a computer. This is possible
using a Fourier expansion of the functions $h$ and $\dot H$ and
truncating to a finite number of Fourier coefficients. One can tune
this finite number until the precision attained in the function
$\tilde f$ is of 14-16 decimal places. Due to the fast convergence of
the Fourier coefficients, this procedure is not very costly. The
computation in a conventional PC takes around 50 hours. 

Once $\tilde f$ is determined at the grid points, in order to extract 
the values of $F^2(2 |u|)$ we proceeded  as follows. 
For a given value of $|u|$ we take many different
values of $w$ from the grid and for each we compute the corresponding 
value of $\epsilon$. For that value  we sum the series up to 
40th order and multiply by the function associated to the change of
variables (Eq.~\ref{eq:metricw}). Finally,  for each value of
$|u|$ we obtain we obtain a collection of numbers for different
values of $\epsilon$ and for different phases of $u$.   
A typical case is displayed in  Fig.~\ref{fig:errscaling}, where we
considered points having $|u|=1$ and situated along the 
diagonal $\mbox{Re}(u)=\mbox{Im}(u)$. The value of $F^2(2 |u|)$
is the limit for $\epsilon$ approaching 1. Unfortunately, the 
larger values of $\epsilon$ are more seriously affected by the 
truncation of the series to 40 orders. This is exemplified by the 
errors attached to the points of the figure. These are obtained 
by subtracting the result obtained for 35 orders to that of 40 
orders. To obtain a more precise determination of the value of 
$F(2 |u|)$ we made  two corrections. First of all, we partially 
corrected for the truncation errors by estimating 
the contribution of the terms from the 41st on. The precise way
in which this is done will be explained later. 
This additional  contribution does not affect significantly points 
for which the error obtained by comparing the  35th and 40th 
order is tiny. On the opposite extreme the contribution becomes 
unreliable when this error becomes very large. This occurs 
typically for values of $\epsilon$ exceeding $0.95$ (areas which 
are more than 20 times the critical area). We will therefore 
omit those points from our analysis. 

To extract a good determination of $F^2(x)$ from our data we face 
the problem of extrapolating to $\epsilon=1$. This implies relating 
the results for a large torus with that of the plane. The torus case 
can be considered a solution of the plane with an infinite number 
of vortices -- two per $l\times l$ cell. As the torus gets larger 
the replica vortices are increasingly far away, so that it is 
reasonable to use the analytic result for the case of a large number 
of asymptotically separated vortices to describe the approach. 
Using the result of Ref.~\cite{Manton:2002wb} we expect an additional 
contribution to the metric proportional to $\sum_r K_0(D_r)$, where
$K_0$ is the  
modified Bessel function of the second kind and $D_r$ is the distance
to the replica vortex $r$. In the limit of large torus sizes and fixed
$|u|$, this predicts a contribution proportional to $K_0(l)$. On the
basis of the asymptotic behaviour of Bessel functions we decided to fit 
our data for each value of $u$ to a formula of the form 
\begin{equation}
\label{eq:fit}
  A-B(1-\epsilon)^{\frac{1}{4}} e^{-\sqrt{8\pi/(1-\epsilon)}}
\end{equation}
For our $|u|=1$ example, the result is displayed by the solid line in 
Fig.\ref{fig:errscaling}, and the fitted parameters are
$B= 145.0(5)$ and $A= F^2(2)= 0.8791(4)$.   The error reflects 
systematics due to changing the range of the fit as well as the
relative weighting of the points. For other phases of $u$ we get
compatible results. For example, for  purely real or imaginary values
of $u$ we obtain $F^2(2)=0.8800(6)$, consistent with rotational
invariance.

\EPSFIGURE[h]{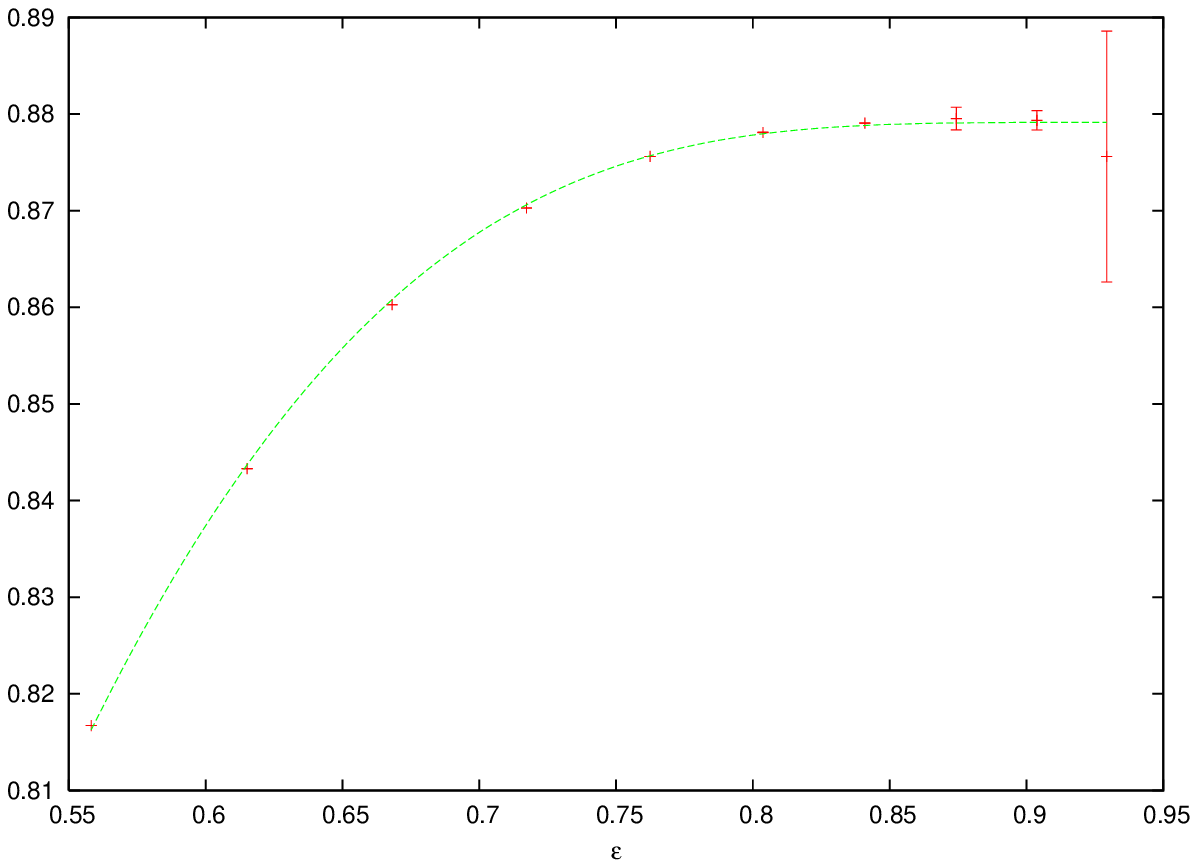}{\label{fig:errscaling}
The points are the values of the left hand side of Eq.~\ref{eq:Flimit} 
for $|u|=1.0$ as a function of $\epsilon$, computed from the first 
40 orders of our expansion.  Error bars represent the size of the 
last 5 terms of the expansion. The solid line is a fit with
Eq.~\ref{eq:fit}.}

Proceeding in the same way for several values of $|u|$ we arrive 
at the points displayed in Fig.~\ref{fig:funcf}. The errors are smaller 
than the size of the points in the figure. In all the range up to
$|u|\ge 2$ the error is smaller than $6\; 10^{-4}$ and for $|u|>0.5$
the relative error is smaller than 1 part in $10^3$.

\EPSFIGURE[h]{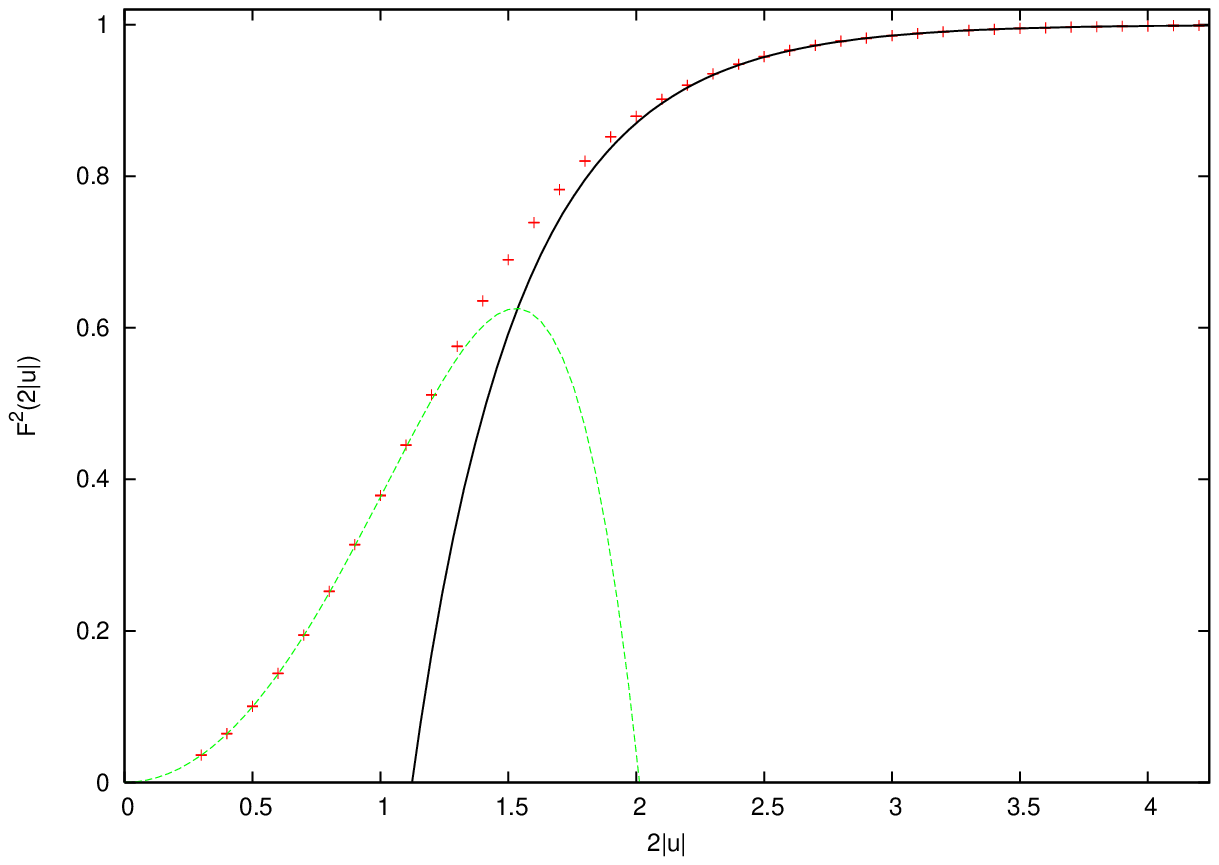}{\label{fig:funcf}The metric computed
  using 40 orders of the Bradlow parameter expansion (red points),
  versus the $|u|\rightarrow 0$ approximation Eq.~\ref{eq:r0} (green
  dashed line) and the asymptotic form of the metric
  Eq.~\ref{eq:metricas} (black solid line).}

It is quite interesting to analyse the implications of the scaling
limit for the coefficients of the $\epsilon$ expansion.
For that purpose it is convenient to write $\epsilon=e^{-\delta}$ and
replace  the limit by $\delta
\longrightarrow 0$. 
The condition that at large volumes the metric tends to that of the
plane implies:
\begin{equation}
  \label{eq:Ae}
  F^2(2|u|) = \lim_{\delta \rightarrow 0} \delta^2
  \frac{2\pi^2Z_2^2}{(1+Z_0^2)^2}|u|^2 \sum_{N=1}^\infty
  {\tilde f}^{(N)}(u\sqrt{ 2 \pi  \delta},\overline u \sqrt{ 2\pi 
    \delta}) e^{-N \delta}
\end{equation}
It is clear that as $\delta$ tends to zero, more terms in the series
become relevant. It is tempting to assume that the infinite sum tends        
to an integral over the variable $y=N\delta$. This would imply the
following behaviour of the coefficients:
\begin{equation}
  \label{eq:CoeBeh}
  \lim_{N \longrightarrow  \infty} {\tilde
    f}_w^{(N)}\left(u\sqrt{\frac{2 \pi y}{N}},\overline u \sqrt{\frac{
    2 \pi
        y}{N}}\right) \frac{\delta}{8} = h(|u|\sqrt{y})  
\end{equation}
In this case the limiting function $h(|u|\sqrt{y})$ would be related 
to $F^2(2|u|)$ as follows:
\begin{equation}
  \label{eq:metricplane}
    F^2(2|u|) =
      \frac{16\pi^2Z_2^2}{(1+Z_0^2)^2}|u|^2 \int_{0}^\infty
        e^{-y} h(|u|\sqrt{y})\, dy
	\end{equation}
which is essentially a Laplace transform. 	

With our results for ${\tilde f}_w^{(N)}$ up to  $N=40$ and the 
set of values of $w$ spanning the aforementioned  $40\times 40$ grid,
we have analysed the  validity of Eq.~\ref{eq:CoeBeh}. 
For that purpose we  plot  in
Fig.~\ref{fig:funch} the coefficients satisfying  $35 \le N \le 40$ and all values 
of $u$ in the range. The smoothness and thinness of the resulting 
curve, approximating $h(x)$, is a test of our scaling hypothesis. 
Errors, signalled by  the spread of the values, grow with $x$. 
For example, including all coefficients $N>20$  would provide little
changes in the thickness of the curve  up to $x\approx 5$.
Beyond, the thickness becomes a sizable fraction of the value.

The  function $h(x)$ for small $x$, as determined from 
our results,  fits nicely to a behaviour of the form
\begin{equation}
  \label{eq:w0}
    h(x) = a - bx^4 + O(x^5)
    \end{equation}
    where $a=0.335(1)$ and $b=0.16(1)$.
This expansion implies that for small values of the argument 
the metric function $F^2(2|u|)$ behaves as:
\begin{equation}
  \label{eq:r0}
  F^2(2 |u|) =
  \frac{16\pi^2Z_2^2}{(1+Z_0^2)^2}
  \left( a|u|^2 - 2b|u|^6 \right) + O(|u|^7)
\end{equation}
The function on the left-hand side of the previous formula is displayed 
in   Fig.~\ref{fig:funcf}. It matches nicely to the behaviour 
of the previously determined  points (in red). For large values of $|u|$
the data also matches with the prediction coming from the asymptotic 
behaviour at large distances (Ref.~\cite{Manton:2002wb}) given by  
\begin{equation}
  \label{eq:metricas}
  F^2(2|u|) \stackrel{|u|\rightarrow \infty}{\longrightarrow} 
  1-8\sqrt{2}K_0(4|u|)
\end{equation}
where $K_0(x)$ is the modified Bessel function of the second kind.

\EPSFIGURE[h]{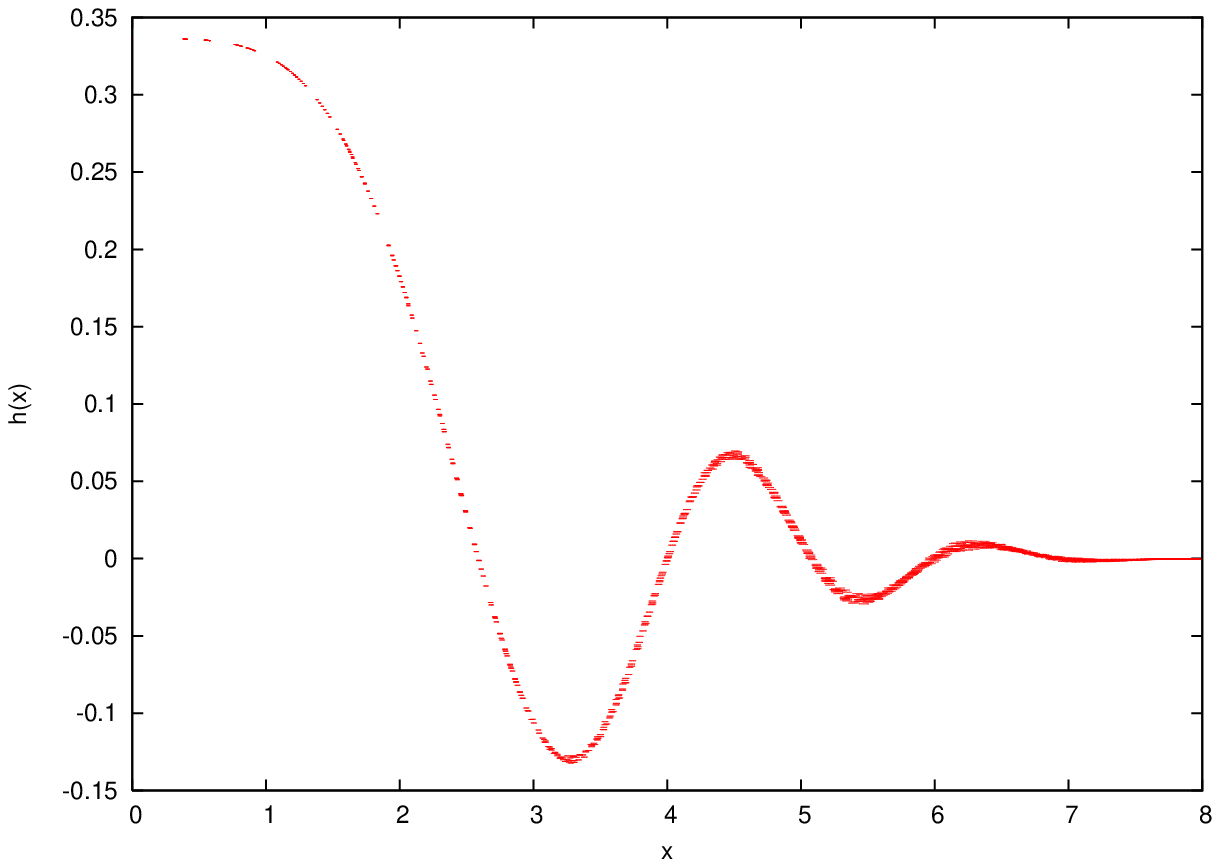}{\label{fig:funch} We plot the left-hand
side of Eq.~\ref{eq:CoeBeh} for $N$ in the range 35-40, and fixed
$x=|u|\sqrt{y}$.}

To conclude we point out that our knowledge of $h(x)$ enables one 
to estimate the error introduced by the truncation in the number 
of terms in the $\epsilon$ expansion. For that one has simply to 
use the expression given in Eq.~\ref{eq:metricplane} with the 
integral restricted to $y>-N\log(\epsilon)$, where $N$ is the maximum 
order computed. We have tested this estimation with our data points 
and $N=40$. As mentioned previously, it improves the results for 
not too large values of $\epsilon$. 

\subsubsection{Scattering of vortices on the plane}

Once the function $F(2 |u|)$ has been obtained, it is trivial to compute
the dynamics of the vortices. The Lagrangian, when written in polar
coordinates $2u = re^{\imath\theta}$, is given by:
\begin{equation}
  T = \pi F^2(r)\left[\dot r^2 + r^2\dot\theta^2\right]
\end{equation}
Since the energy is conserved, and we have rotational invariance
($p_\theta$ is also conserved), the system is integrable, and we have 
\begin{equation}
  \theta_f = \pi - \int^\infty_{r_f}
  \frac{dr}{r\sqrt{\frac{r^2}{b^2}F^2(2r) - 1}} 
\end{equation}
where $b^2=\frac{\pi p_\theta^2}{4T}$ is the \emph{impact parameter}. In
particular, if we call $r_m$ the minimum distance between one vortex
and the centre of mass during the trajectory, the scattering angle is
given by 
\begin{equation}
  \theta_{sc} = \pi - 2\int_{r_m}^{\infty}
  \frac{dr}{r\sqrt{\frac{r^2}{b^2}F^2(r) - 1}} 
\end{equation}

Using our numerical data of
Fig.~\ref{fig:funcf} we can calculate these integrals. For values of
$r > 5$ we will use the asymptotic form of the metric, and for 
$r<0.5$ we can use our approximation Eq.~(\ref{eq:r0}). For the
intermediate values we will use the cubic spline interpolating
polynomial. In the Fig.~\ref{fig:scatt} we have some examples of
trajectories for different values of the impact parameter.  

\EPSFIGURE[h]{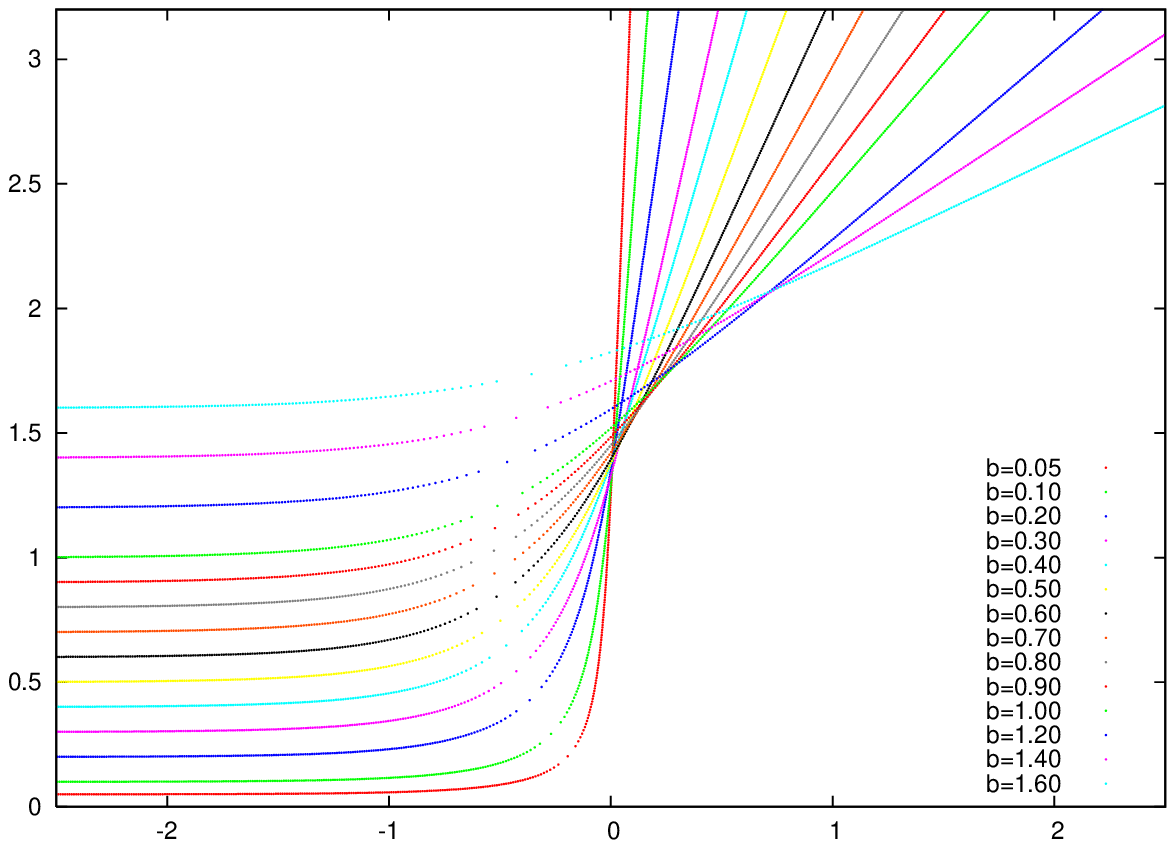}{\label{fig:scatt}Scattering trajectories
  of two vortices.} 

We can also plot the scattering angle versus the impact parameter. The
results are shown in Fig.~\ref{fig:angle}. Errors were estimated 
in the following way. We generate a random set  of values  of $F(2|u|)$
assuming a Gaussian distribution around the central value and standard 
deviation fixed by the error. For each realization we 
compute the cubic spline interpolation polynomial and using 
it and the small and large approximation functions we compute the 
scattering angle for each impact parameter. The final errors are
estimated on the basis of 1000 realizations. They are 
of the order of 1 part in $10^4$ or smaller.

\EPSFIGURE[h]{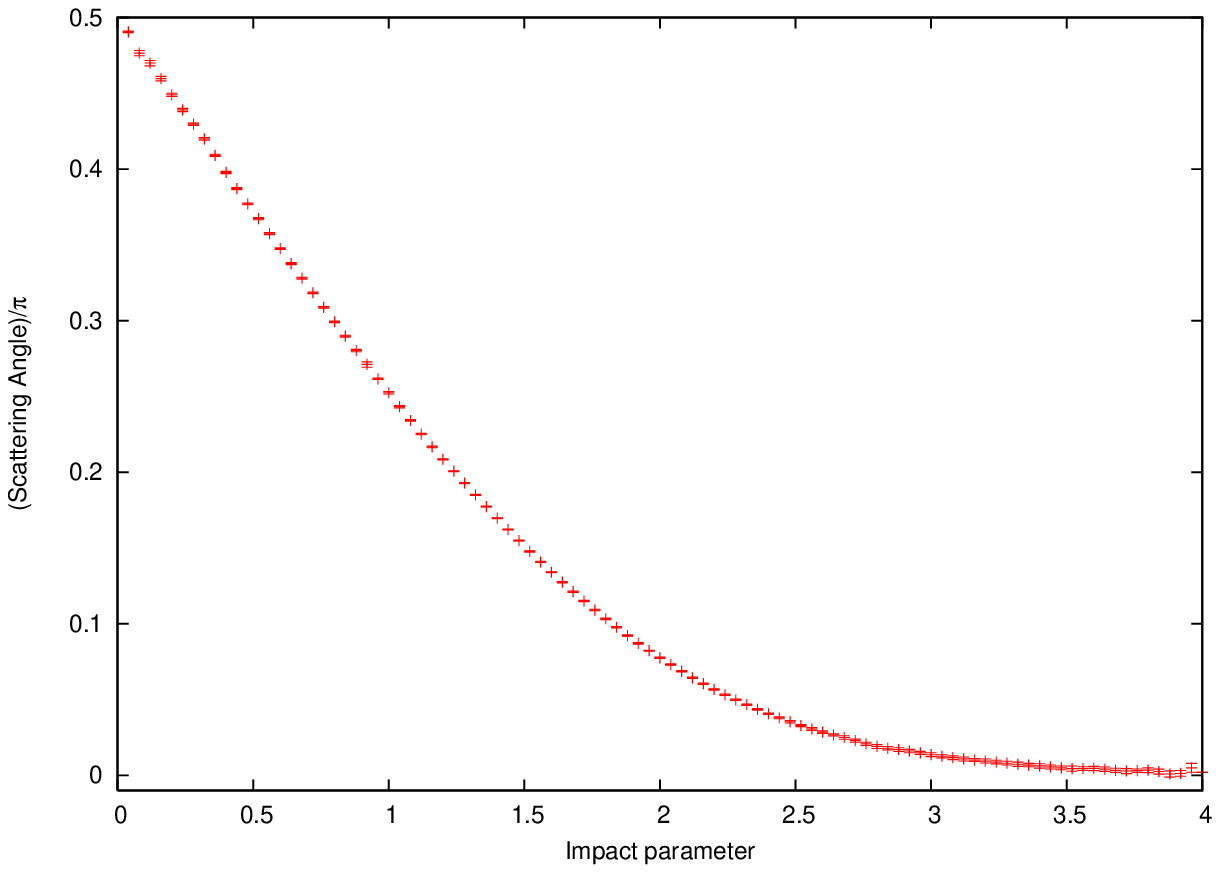}{\label{fig:angle}Scattering Angle
  vs. impact parameter.}

\section{Conclusions}
In this paper we have shown how the Bradlow parameter expansion
can be used to compute the metric of the moduli space  of
solutions of the Abelian Higgs model on the Torus.
This extends our previous work~\cite{Gonzalez-Arroyo:2004xu} where 
this expansion was introduced as a tool to construct 
the classical vortex solutions on the  Torus. More precisely, the 
expansion is performed in the variable $\epsilon=1-2f$, where $f$ is
the average magnetic field. To any finite order in the expansion 
the metric has a known form depending on  a small number of real parameters. 
We give their analytic expression up to second order for arbitrary 
number of vortices $q$. The leading term, giving  the Bradlow limit
result, is proportional to the Fubini-Study metric. This coincides 
with results obtained for the two-sphere~\cite{Baptista:2002kb}. The second order 
result depends on $q^2$ constants, which are given in the form of rapidly 
decreasing double infinite sums. Although increasingly complicated 
it is not too hard to go beyond this order. Rather than attempting this
for  the general vortex case, we have concentrated on the two-vortex 
case. We have given the form of the metric up to 5th order and the
numerical value of their 15 real  coefficients with 14-16 significant digits 
(machine double precision). Furthermore, it is possible to go much
beyond this order in the computation of the metric at specific points 
of the moduli space by means of a computer. Essentially, the
implementation is based on the truncation of the infinite sums to
finite ones. In this way we have gone up to 40th order with fairly
limited computational effort. Furthermore, the fast convergence of the 
sums stills maintains the double machine precision accuracy in the 
determination of the metric. Having so many terms in the expansion 
has allowed us to extrapolate our results to the infinite area case
(vortices on the plane) with quite impressive results. The resulting
metric is consistent with all the previously known results, both 
analytical in certain limits and numerical. The result have fairly small
and controlled errors,  arising from the extrapolation to infinite
volume and the truncation in the expansion. 

The computation of the metric allows the study of vortex scattering,
known to be a good approximation up to relatively high vortex
velocities. The presence of the torus can be used to mimic the
behaviour of vortices in a dense environment. A natural question that 
poses itself is the possibility that two-vortex dynamics on the torus 
is integrable. This is actually the case for both small (close to
critical) torus sizes  as for infinitely large  ones (the plane). 
We have analysed  Poincare maps as a first step in this study.
Furthermore, we emphasise that, as in our previous work,  
our method has been found to be a  competitive
tool to study the properties of vortices  on the plane. It has 
important advantages over other methods: its semi-analytical nature 
and the possibility of dealing with an arbitrary number of vortices.
In this work, for example,  we have given  the curves (with errors)
that describe the functional dependence of the two-vortex  scattering angle 
with the impact parameter. 

To conclude, we want to mention that the Bradlow expansion method is 
generalisable to many other cases and theories. For example, it can
be applied to  any other  two-dimensional compact manifold.
Furthermore, a similar expansion seems to apply in other cases where 
a Bradlow parameter can be introduced~\cite{Bradlow:1990ir}, such as abelian
and non-abelian gauge-Higgs theories in K\"ahler manifolds of arbitrary 
dimensionality. The methodology can be extended beyond. For example, 
a related expansion, is known to exist in the study of self-dual 
configurations on the 4 dimensional torus~\cite{GarciaPerez:2000yt}.
Recently~\cite{Forgacs:2005ct,Lozano:2006xn}, 
it has also been applied to the study of abelian vortices in a 
non commutative torus. In all these theories,  apart from the
determination of the shape 
of the solutions and the form of the metric of the moduli space, other 
problems can be attacked. In particular, one can study  the 
eigenfunctions and eigenvalues of the Dirac operator in a vortex 
background and also compute quantum corrections to the vortex.
These problems are currently under study by the present authors.

\acknowledgments
We acknowledge financial support from the Comunidad Aut\'onoma
de Madrid under the program PRICyT-CM-5 S-0505/ESP-0346, and 
from the Spanish Ministry of Science and Education  under
contracts FPA2003-03801 and FPA2003-04597. A. R. wants to thank 
N. Manton and the research group  at DAMPT (Cambridge) for their
hospitality during the last stages of this work, and S. Krusch for
his useful comments on the manuscript.

\appendix

\section{Apendix:}
\label{ap:order2}
To lowest order we have that $h^{(0)}=\ln (||c||) $
and $\dot{H}^{(0)}=\frac{\sum_i \overline{c}_i \dot{c}_i}{||c||^2}$ 
are constant. The symbol $||c||^2$ stands for $\sum_i |c_i|^2$.
Hence, the lowest order 
metric is given by 
\begin{equation}
g^{(1)}_{ij} = \frac{{\cal A}  }{\pi ||c||^2} \left(
    \delta_{ij} - \frac{\overline c_i c_j}{||c||^2}
      \right)     
\end{equation}
The matrix inside parenthesis, which we will call ${\mathcal P}$,  is the 
orthogonal projector to the vector $c_i$. It gives precisely the 
Fubini-Study metric written in homogeneous coordinates. 

To the next order the metric, written in matrix notation, adopts the form:
\begin{equation}
g^{(2)}= \frac{{\cal A}  }{\pi ||c||^2}\   {\mathcal P} \left(
-{\mathbf I} \frac{\sum_{i j} \overline{c}_i \overline{S}_{ i j} c_j
}{||c||^2} + \overline{S} + \widetilde{S} 
\right) {\mathcal P}
\end{equation}
where 
\begin{equation}
\overline{S}_{ i j} = \frac{\overline{c}_l c_k }{{\mathcal A} ||c||^2}
\int d^2 x \overline{\chi}_i \chi_j  \Delta^{-1} (\overline{\chi}_l
\chi_k)'  
\end{equation}
and 
\begin{equation}
\widetilde{S}_{ i j} = \frac{\overline{c}_l c_k }{{\mathcal A} ||c||^2}
\int d^2 x
\overline{\chi}_i \chi_k  \Delta^{-1} (\overline{\chi}_l \chi_j)'
\end{equation}
The prime denotes the removal of the constant Fourier mode,
making the inverse laplacian $\Delta^{-1}$ well-defined. 

The computation of the integrals can be explicitly carried over by
using Fourier transforms. For that purpose we introduce the quantities:
\begin{equation}
  L^{ij}(\vec n) = \frac{1}{\mathcal A}\int_{\mathbb T^2}d^2x\,
  \overline{\chi}_i e^{-2\pi \imath n_i \cdot\frac{x_i}{l_i}} \chi_j
\end{equation}
which are just  the Fourier modes of the product
$\overline{\chi}_i \chi_j$. Now we remove the constant mode ($\vec n =0$)
and denote the result as $L'^{ij}(\vec n)$.
Then the integral entering in the definition of $\overline{S}$ 
and $\widetilde{S}$ is given by 
\begin{equation}
\label{eq:sijkl}
 S_{ijkl} \equiv -\frac{1}{\mathcal A} \int_{\mathbb T^2}  d^2 x
 \overline{\chi}_i \chi_j  \Delta^{-1} (\overline{\chi}_l \chi_k)'=
 \sum_{\vec n}\frac{L'^{ij}(\vec n)L'^{lk}(-\vec n)}{\xi(\vec n)}
\end{equation}
where 
\begin{equation}
  \xi(\vec n) = \frac{\pi\tau}{q}\left[n_1^2 + \frac{n_2^2}{\tau^2}\right]
\end{equation}
and $\tau = l_2/l_1$. The explicit value of the coefficients depends on
the choice of basis. For the choice given earlier $\chi_{s+1}=\Psi_{0 s}$
defined in equation $(A.24)$ of our previous
paper~\cite{Gonzalez-Arroyo:2004xu}, the coefficients are given (as a
particular case) of Eq.~(A.30) of that paper. The matrix ${\mathbf
  L}(\vec{n})$ becomes:
  \begin{equation}
{\mathbf L}(n)={\mathbf \Gamma}(\vec{n})   e^{-\frac{\xi(\vec{n})}{2}}
\end{equation}
where the matrices ${\mathbf \Gamma}(\vec{n})$
are defined as 
\begin{equation}
 {\mathbf \Gamma}(\vec{n}) = P^{n_1} Q^{n_2} e^{-\imath \pi \frac{n_1 n_2}{q}}
\end{equation}
where $P$ and $Q$ are `t Hooft $q\times q$ matrices ($P_{i j}=
\delta_{j i+1}$, $Q=\mbox{\rm diag}(\exp{ 2 \pi i s/q})$). 
The matrices ${\mathbf \Gamma}(\vec{n})$ are unitary and satisfy 
${\mathbf \Gamma}^\dagger(\vec{n})={\mathbf \Gamma}(-\vec{n})$.
From here  one realises that, the coefficients 
$S_{ijkl}$ can be expressed in terms of $q^2$ constants $K(\vec{s})$ 
as follows:
\begin{equation}
  S_{ijkl} = \sum_{\vec{s}} K(\vec{s})\,  \Gamma_{ij}(\vec{s})
 \,  \Gamma_{k l}(\vec{-s})
\end{equation}
where $\vec{s}$ is a  2 component vector with components
$s_i\in \{0,\ldots, q-1\} $. From Eq.~\ref{eq:sijkl} one derives:
\begin{equation}
  K(\vec{s})=\sum_{\vec{k}\in Z^2} \frac{e^{-\xi(\vec{s}+q
      \vec{k})}}{\xi(\vec{s}+q \vec{k})}=\int_1^\infty d\alpha
  \left(\thetachar{s_1/q}{0}(0|i q\alpha\tau)
 \;   \thetachar{s_2/q}{0}(0;i q\alpha/\tau)-1\right) 
\end{equation}
The terms  in the sum decay  exponentially with $k^2$,
so that a few terms in the infinite sums over $\vec k$ suffice to obtain 
a very high accuracy. 

The matrices ${\mathbf \Gamma}(\vec{s})$ are a basis of the set of 
$q\times q$ matrices. Hence, it is possible to express the metric tensor 
to this order as a linear combination of these matrices. The coefficients
are functions of the following quadratic forms in $c_i/||c||$:
\begin{equation}
  P(\vec{s})=\frac{\overline c {\mathbf \Gamma}(-\vec{s}) c}{||c||^2} 
\end{equation}
These are not independent, and satisfy  certain relations. In
particular one has $P(\vec 0)=1$ and 
\begin{equation}
\sum_{\vec{s}\neq \vec 0} |P(\vec{s})|^2=q-1
\end{equation}

As an example, we give the explicit expression for the trace of
$g^{(2)}$:
\begin{equation}
  ||c||^2  {\rm Tr}(g^{(2)})= \frac{{\cal A}  }{\pi }\
  \sum_{\vec{s}\neq \vec 0} K(\vec{s}) ((q+1) |P(\vec{s})|^2-1)
\end{equation}
From here it is trivial to reproduce the $q=2$ case given in
section~\ref{sc:bradlow}. Notice that $(P(1,0),P(0,1),P(1,1))$ is a
3-component unit vector whose expression in terms of spherical coordinates
coincides with our definition of $\theta$ and $\varphi$ in
Eq.~\ref{eq:basistp}. For the symmetric $l_1=l_2$ case we obtain 
\begin{equation}
  \tilde f^{(2)}= \frac{{\cal A}  }{\pi }\ (K(1,0)-K(1,1)) (1-3
  |P(1,1)|^2)
\end{equation}
where $P(1,1)=\cos\theta$. This reproduces the result given in 
Table~\ref{tab:metric}. Notice that the result is proportional to the 
second Legendre polynomial in $\cos{\theta}$ as follows from the 
area condition. 

Going over to higher orders is tedious but presents essentially no new 
difficulties. We have obtained the analytic result to third order for the 
$q=2$, $\tau=1$ case, matching the value given in Table~\ref{tab:metric}.

\bibliography{}

\end{document}